\documentclass[twocolumn]{aastex631}

\usepackage{natbib}
\usepackage{soul}
\usepackage{hyperref}
\usepackage{graphicx}

\providecommand{\e}[1]{\ensuremath{\times 10^{#1}}}	            
\newcommand{\HII}{\ion{H}{2}}		        	
\providecommand{\HI}{\ion{H}{1}}               		        	
\providecommand{\HA}{H$\alpha$}		                			
\providecommand{\R}{$R_{23}$}		                			

\received{}
\revised{}
\accepted{}
\shorttitle{NGC~6822 Spitzer I YSOs}
\shortauthors{Lenki\'{c} et al.}
\graphicspath{{./}{figures/}}

\begin{document}

\title{A JWST/MIRI and NIRCam Analysis of the Young Stellar Object Population in the Spitzer I Region of NGC 6822}

\correspondingauthor{Laura Lenki\'{c}}
\email{laura.lenkic@gmail.com}

\author[0000-0003-4023-8657]{Laura Lenki\'{c}}
\affil{Stratospheric Observatory for Infrared Astronomy, NASA Ames Research Center, Mail Stop 204-14, Moffett Field, CA 94035, USA}
\affil{Jet Propulsion Laboratory, California Institute of Technology, 4800 Oak Grove Dr., Pasadena, CA 91109, USA}

\author[0000-0002-7512-1662]{Conor Nally}
\affil{Institute for Astronomy, University of Edinburgh, Blackford Hill, Edinburgh, EH9 3HJ, UK}

\author[0000-0003-4870-5547]{Olivia C.\ Jones}
\affil{UK Astronomy Technology Centre, Royal Observatory, Blackford Hill, Edinburgh, EH9 3HJ, UK}

\author[0000-0003-4850-9589]{Martha L.\ Boyer}
\affil{Space Telescope Science Institute, 3700 San Martin Drive, Baltimore, MD 21218, USA}

\author[0000-0001-6872-2358]{Patrick J.\ Kavanagh}
\affil{Department of Experimental Physics, Maynooth University-National University of Ireland Maynooth, Maynooth, Co Kildare, Ireland}

\author[0000-0002-2667-1676]{Nolan Habel}
\affil{Stratospheric Observatory for Infrared Astronomy, NASA Ames Research Center, Mail Stop 204-14, Moffett Field, CA 94035, USA}
\affil{Jet Propulsion Laboratory, California Institute of Technology, 4800 Oak Grove Dr., Pasadena, CA 91109, USA}

\author[0000-0001-6576-6339]{Omnarayani Nayak}
\affil{Space Telescope Science Institute, 3700 San Martin Drive, Baltimore, MD 21218, USA}
\affil{NASA Goddard Space Flight Center, 8800 Greenbelt Road, Greenbelt, MD, USA}

\author[0000-0002-2954-8622]{Alec S.\ Hirschauer}
\affil{Space Telescope Science Institute, 3700 San Martin Drive, Baltimore, MD 21218, USA}

\author[0000-0002-0522-3743]{Margaret Meixner}
\affil{Stratospheric Observatory for Infrared Astronomy, NASA Ames Research Center, Mail Stop 204-14, Moffett Field, CA 94035, USA}
\affil{Jet Propulsion Laboratory, California Institute of Technology, 4800 Oak Grove Dr., Pasadena, CA 91109, USA}
\affil{Space Telescope Science Institute, 3700 San Martin Drive, Baltimore, MD 21218, USA}
\affil{Department of Physics \& Astronomy, Johns Hopkins University, 3400 N.\ Charles St., Baltimore, MD 21218, USA}

\author[0000-0001-9855-8261]{B.\ A.\ Sargent}
\affil{Space Telescope Science Institute, 3700 San Martin Drive, Baltimore, MD 21218, USA}
\affil{Department of Physics \& Astronomy, Johns Hopkins University, 3400 N.\ Charles St., Baltimore, MD 21218, USA}

\author[0000-0001-7380-3144]{Tea Temim}
\affil{Department of Astrophysical Sciences, Princeton University, Princeton, NJ 08544, USA}



\begin{abstract}
\noindent We present an imaging survey of the Spitzer~I star-forming region in NGC~6822 conducted with the NIRCam and MIRI instruments onboard \emph{JWST}. Located at a distance of 490~kpc, NGC~6822 is the nearest non-interacting low-metallicity ($\sim$0.2~$Z_{\odot}$) dwarf galaxy. It hosts some of the brightest known \HII{} regions in the local universe, including recently discovered sites of highly-embedded active star formation. Of these, Spitzer~I is the youngest and most active, and houses 90 color-selected candidate young stellar objects (YSOs) identified from \textit{Spitzer Space Telescope} observations. We revisit the YSO population of Spitzer~I with these new \emph{JWST} observations. By analyzing color-magnitude diagrams (CMDs) constructed with NIRCam and MIRI data, we establish color selection criteria and construct spectral energy distributions (SEDs) to identify candidate YSOs and characterize the full population of young stars, from the most embedded phase to the more evolved stages. In this way, we have identified 140 YSOs in Spitzer~I. Comparing to previous \textit{Spitzer} studies of the NGC~6822 YSO population, we find that the YSOs we identify are fainter and less massive, indicating that the improved resolution of \emph{JWST} allows us to resolve previously blended sources into multiple objects.
\end{abstract}




\section{Introduction} 
\label{sec:intro}
NGC~6822 is an isolated dwarf irregular galaxy located in the Local Group at a distance of $490 \pm 40$~kpc \citep{sibbons15}, with a metallicity comparable to that of the Small Magellanic Cloud \citep[$\sim$0.2~Z$_{\odot}$;][]{garcia16}. It is characterized by a central bar, oriented in a north-south direction, that contains most of the young stellar population of the galaxy \citep{schruba17}, an \HI{} disk that extends well beyond the optical extent \citep{deblok00,deblok06}, and several prominent \HII{} regions and OB associations \citep{efremova11,rubin16}. These \HII{} regions are among the most massive and brightest known in the local universe \citep{hubble25} and span a range of evolutionary stages \citep{schruba17,jones19}. 

The best-known \HII{} regions (i.e., Hubble I/III, V, and X along the northern part of the galaxy, and Hubble IV in the south) have been studied in detail through numerous multi-wavelength observations: ultraviolet \citep[UV;][]{efremova11}, H$\alpha$ \citep{kennicutt79,hodge88,massey07}, near- and mid-infrared \citep[IR;][]{cannon06}, far-IR \citep{gallagher91,israel96}, and molecular gas \citep{israel03,schruba17}. These studies show that the Hubble regions host massive young stellar objects (YSOs) and are actively forming stars, with an average star formation rate (SFR) of $1.0 \times 10^{-2}$~M$_{\odot}$\,yr$^{-1}$ over the past 10~Myr.
In addition, these star-forming regions appear to exist at varying evolutionary stages from one another, based on differing measured SFRs and demographics of young and pre-main sequence (PMS) stars \citep{jones19, kinson21}.  These star-formation properties make NGC~6822 a useful system for understanding extragalactic activity and evolution in the early universe (e.g., \citealp{bib:Lee2005, bib:Hunter2007, rubin16}).

Observations in the IR provide a window into the earliest phases of formation for super star clusters (SSCs), which possess $>$10$^{5}$~M$_{\odot}$ of stars and are thought to be precursors to globular clusters.
The recent discovery of the proto-SSC candidate Spitzer I, as well as the embedded star-forming regions Spitzer II and III, provide further evidence for the existence of new massive star formation in NGC~6822 \citep{jones19, hirschauer20}.
Compared to the optically-identified Hubble regions, the IR-detected star-forming clusters were found to be younger, with many more early-stage YSOs as determined by spectral energy distribution (SED) analysis.
Furthermore, because they possess a higher level of IR flux than \HA, their star-formation activity is expected to be increasing \citep{jones19}.
These characteristics place Spitzer I among other SSCs, including  Westerlund I in the Milky Way, which possesses at least 200 massive stars \citep{bib:Clark2005}, and R136 in the 30 Doradus region of the Large Magellanic Cloud (LMC), home to dozens of O3 stars, including several with masses $\gtrsim$120~M$_{\odot}$, within only two parsecs from the center of the cluster \citep{bib:Hunter1995}.
With \emph{JWST} observations of these SSCs, as well as a forthcoming JWST program (Nayak et al., in prep.) to observe the proto-SSC candidate H72.97-69.39 in the N79 region of the LMC \citep{ochsendorf17, nayak19, bib:Andersen2021}, it becomes possible to perform comparative analyses of the characteristics of SSCs across a range of evolutionary stage, environment, and level of chemical enrichment.

Detailed study of the populations of young stars in these regions will allow for a more complete understanding of the physical mechanisms governing star formation in environments typical of the early universe.
Photometry in the IR traces the beginnings of stellar lifetimes:
YSOs are birthed in active star-forming regions and exhibit strong IR-excess as light is absorbed and re-emitted by cool, dusty envelopes and accretion disks.
With \emph{Spitzer} photometry, the massive stars ($\gtrsim$8 $M_{\odot}$) and star clusters that are accessible account for only one in every ten thousand, and because YSOs progress rapidly through their evolutionary stages, observations are quite rare.
\citet{jones19} determined a population of massive YSOs in Spitzer~I of 90 sources, nearly twice as many as were found in the next-most massive star-forming regions Hubble IV and V (53 sources each), while \citet{kinson21} found 139 candidate YSOs within the same region by applying machine learning (ML) techniques to near-IR colors and far-IR surface brightness measurements. Because it is impossible to conduct such resolved studies for the earliest galaxies, nearby regions with comparable metallicities offer the best laboratories for study.
Harnessing the high sensitivity and resolution afforded by \emph{JWST} \citep{gardner23,rigby23}, our observing program provides a census of young stars down to a stellar mass of $\sim$2 $M_{\odot}$, allowing for a more complete understanding of star formation in environments analogous to the early universe.

In this work, we present \emph{JWST} observations of the central stellar bar of NGC~6822 taken with the Near-Infrared Camera \citep[NIRCam;][]{rieke05, rieke23} and Mid-Infrared Instrument \citep[MIRI;][]{rieke15, wright2023}, which provide unprecedented detail of the stellar populations and interstellar medium (ISM) from the near- to mid-IR. We will focus on the young stellar population of the Spitzer~I region, while an accompanying paper will discuss the NGC~6822 program as a whole, including the parallel fields, and the evolved star population \citep[][]{nally2023}. In \S\ref{sec:obsprog} we describe the observing program, data processing, and photometry, in \S\ref{sec:prelimresults} we present color images of Spitzer~I and our methods for selecting YSOs, in \S\ref{sec:discussion} we compare our results to previous estimates of the YSO population in Spitzer~I and YSO studies in other nearby low-metallicity star-forming regions, and finally in \S\ref{sec:summary} we summarize our findings and discuss future work on studying NGC~6822 and the Spitzer~I candidate proto-SSC. 


\section{Observations and Data Processing} 
\label{sec:obsprog}
\subsection{Observations}
\begin{figure*}
    \centering
    \includegraphics[width=0.49\textwidth]{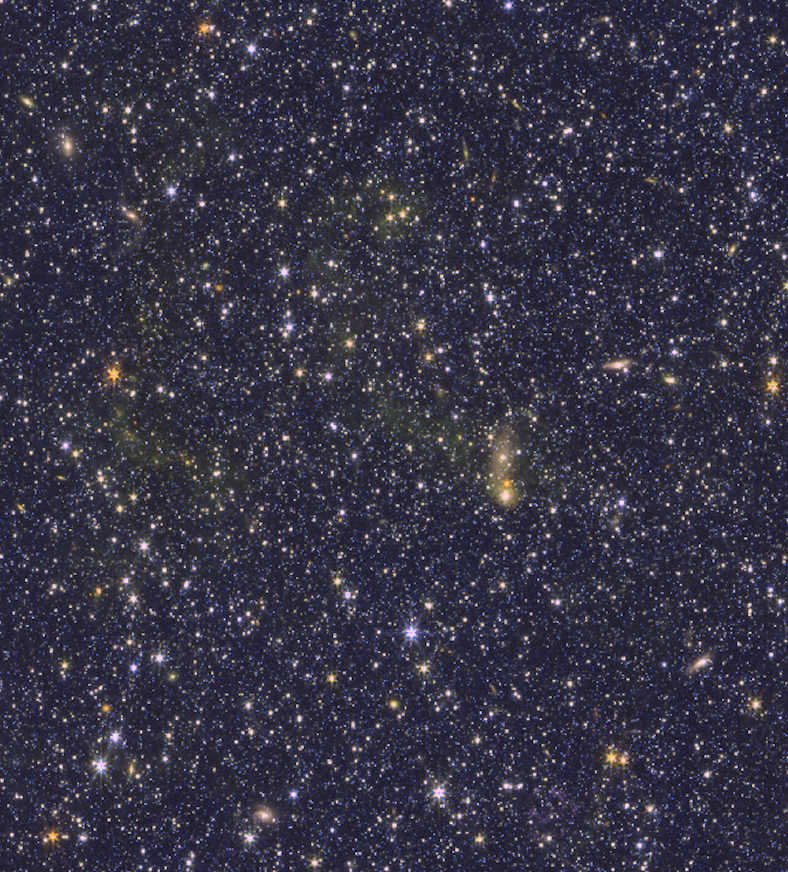}
    \includegraphics[width=0.49\textwidth]{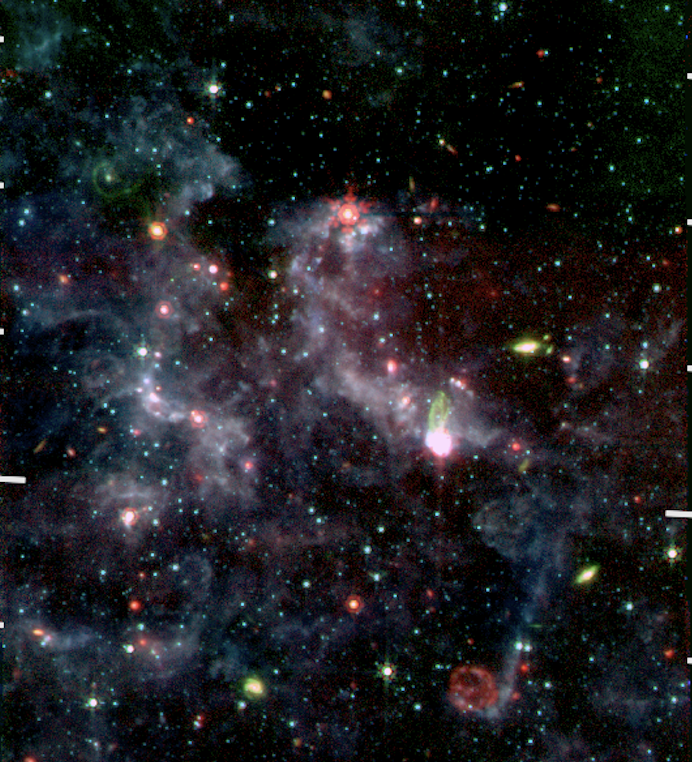}
    \caption{\textit{Left}: Three-color image of Spitzer~I combining the NIRCam F115W (blue), F356W (green), and F444W (red) filters. \textit{Right:} Three-color image of Spitzer~I combining the MIRI F770W (blue), F1000W (green), and F2100W (red) filters. NIRCam and MIRI provide very different views of this active star-forming region. NIRCam reveals a densely populated field of stars with faint emission in the F356W band (green), which may be tracing the 3.3~\micron{} PAH feature. In contrast, MIRI reveals a network of dusty, filamentary structures including the F770W filter (blue) which traces the 7.7~\micron{} PAH feature.}
    \label{fig:color_imgs}
\end{figure*}

We present here observations of NGC~6822 taken with the NIRCam and MIRI instruments onboard \emph{JWST} as part of the GTO program ID:1234 (PI: M.\ Meixner).
Observations with NIRCam were taken on 2022 September 4 for a cumulative $\sim$4.80 hours with the F115W and F200W short wavelength filters and F356W and F444W long wavelength filters using both the A and B modules. These filters were selected in an effort to match existing stellar populations work employing standard Johnson \emph{J} and \emph{K$_{\rm s}$} as well as \emph{Spitzer} IRAC 3.6 and 4.5~$\mu$m filters, respectively. The observations were taken with the \textsc{fullbox 4tight} primary dither pattern and a 10.0\% overlap in rows and columns to ensure no gaps are present in the 2$\times$1 NIRCam mosaic. We also employed a three-point standard subpixel dither pattern to sample the point-spread function (PSF). We used the \textsc{bright2} readout pattern, one integration per exposure with seven groups per integration, and 12 dithers in an effort to optimize S/N (following known best practices at the time of planning) for a total of 1,803~sec integration time per filter. The NIRCam mosaics are centered at RA$\,=\,$19:44:56.1990, Dec$\,=\,-$14:47:51.29 and cover an area of ${\sim} 6.0 \times 4.3$~arcmin$^{2}$.

We processed the NIRCam observations using the \emph{JWST} pipeline \citep[version 1.9.6;][]{bushouse_howard_2023_7714020} with the \texttt{jwst\_1075.pmap} context. This makes use of the latest on-sky derived photometric zero-points \citep{gordon22, boyer22} that were implemented in 2022 October as part of \texttt{jwst\_0989.pmap} in the Operational Pipeline Calibration Reference Data System (CRDS) version 11.16.20. We applied an additional correction to the level 2 calibrated frames from \texttt{Image2Pipeline} (*\_cal.fits files) in order to correct for striping from 1/f noise using the python routine \texttt{image1overf.py}\footnote{\protect\url{https://github.com/chriswillott/jwst.git}}. We then combined these corrected level 2 frames with \texttt{Image3Pipeline}, creating mosaics for each of our four filters that are aligned to the GAIA-DR3 WCS frame using the {\em JWST/Hubble} Alignment Tool \citep[JHAT;][]{jhat}. For additional details pertaining to the NIRCam image processing, see the overview paper \citep{nally2023}.

Our MIRI observations consist of a $6\times1$ mosaic, where five of the tiles were observed on 2022 September 4 and the final tile was observed on 2022 September 15 for a total of $\sim$15.12 hours. Observations were taken using the F770W, F1000W, F1500W, and F2100W filters, selected based on predictions of fluxes for relevant stellar populations by \citet{Jones2017}, with the \textsc{cycling} dither pattern using four positions. We used the \textsc{fastr1} readout pattern with numbers for groups per integration and integrations per exposure determined for each filter using the exposure time calculator (ETC) and applying recommended best practices for maximizing S/N while avoiding saturation \citep[see][for additional details]{nally2023}. The MIRI mosaics are centered at RA$\,=\,$19:44:58.0949, Dec$\,=\,$$-$14:48:20.620 and cover an area of ${\sim} 7.5 \times 1.9$~arcmin$^{2}$. 

We created calibrated MIRI mosaics using {\em JWST} pipeline version 1.9.5 with CRDS version 11.16.21 and context \texttt{jwst\_1084.pmap}. Each of the raw MIRI ramp files was processed through \texttt{Dectector1Pipeline} and the output through \texttt{Image2Pipeline} with default parameters to produce flux-calibrated dither images across all filters. We used the \texttt{tweakreg} step on the resulting files to determine and apply an astrometric correction to all dithers in each mosaic tile individually to align to {\em Gaia} DR2. Since no dedicated background observations were taken, we generated backgrounds from Visits 001 and 005 of Observation 007, i.e.,~the two tiles at the mosaic edges, as these are least affected by real diffuse emission in NGC~6822. A detailed description of the background treatment is presented in \citet{nally2023}. We constructed mosaics from the background subtracted images for each filter using the \texttt{Image3Pipeline} with the \texttt{tweakreg} step switched off as the astrometric correction was already applied. We present in Figure \ref{fig:color_imgs} multi-color images of the NIRCam and MIRI mosaics zoomed onto the Spitzer I region.

\subsection{Source Detection and Photometry}
We perform source detection and photometry using the \textsc{starbugii} tool \citep{nally22}, which has been developed and optimized for \emph{JWST} NIRCam and MIRI observations in crowded fields with complex backgrounds. We perform source detection using \texttt{starbug2 $--$detect} on the individual \textit{Gaia}-aligned level 2 frames. For each wavelength band, we match detections between dithers. In NIRCam, where our dither mode results in three or more overlapping exposures for a given pointing, we require that a detection be matched in two or more exposures to be included in our catalog. In MIRI, because our dither mode results in four or more overlapping exposures for a given pointing, we require sources to be detected in at least three frames. Sources not meeting these thresholds are discarded, thus cleaning the catalogue of most detector artifacts, stray cosmic rays, and snowballs. After source detection is completed, we employ a mix of aperture and PSF photometry to build a photometric catalog of sources in our observations. For both instruments, we first perform single-frame aperture photometry using the \textsc{starbugii} tool. To limit the effects of crowding, we take a consistent tight aperture radius of 1.5 pixels and a sky annulus radius $3-4.5$ pixels across all NIRCam bands. For MIRI, we adjust this aperture radius to 2.5 pixels for F770W and F1000W, and to 3.0 pixels for the F1500W and F2100W bands, and calculate backgrounds using annuli of $4-5.5$ and $4.5-6$ pixels respectively. We then apply the latest, at the time of processing, aperture corrections from CRDS (\texttt{jwst\_miri\_apcorr\_0005.fits} and \texttt{jwst\_nircam\_apcorr\_0004.fits}). Background galaxy contaminants are removed with cuts on the geometric parameters of each source generated by \textsc{starbugii} using limits of \texttt{sharpness$\;=0.4-0.9$} and \texttt{roundness$\;\leq |1.0|$} for NIRCam.

For our NIRCam observations, we additionally perform PSF-photometry, again using \textsc{starbugii}. This routine creates a background image by masking sources detected in the aperture photometry step. The resulting image represents diffuse nebulous emission and is subtracted from the original frames. Using {\sc webbpsf}~\citep{webbpsf2014} version 1.1.1, we generate a 5\arcsec{} radius PSF for each subarray of the NIRCam detectors. We then run \textsc{starbugii}'s PSF fitting routine on the clean, background subtracted image, using the source positions detected during the aperture photometry. This routine fits both a flux and position, allowing the centroid to be kept free within 0.1\arcsec{} of the initial position. In cases where this is exceeded, the flux is refit using the coordinates of the initial detection. We use our aperture photometry catalog to apply a necessary zero-point correction to the PSF photometry measurements. We select sources in our aperture catalog of intermediate magnitude with low uncertainties (${<}5\%$) and match these aperture detected sources to their PSF counterparts. This process is expounded upon in detail in \citep{nally22} and \citep{nally2023}. We then derive the instrumental zero-point from the median photometric difference in both catalogs and apply these corrections to each NIRCam filter. Current MIRI PSFs simulated from {\sc webbpsf} lack the cruciform component which results in poor PSF fitting, thus in this work we build our catalog using aperture photometry only for MIRI bands.

Following the procedures described above, we create independent NIRCam and MIRI catalogs before combining them together. The individual NIRCam filter catalogs are matched in order from the shortest wavelength (F115W) to the longest (F444W) with an increasing separation threshold. The position of a given source is taken from the shortest wavelength it appears in. New sources are appended to the end of the catalog, such that it grows as red sources with rising SEDs are detected at longer wavelengths, and are thus not neglected for not being detected at shorter wavelengths. We use a separation threshold of 0.06\arcsec{} when matching between F115W and F200W and 0.1\arcsec{} between F356W and F444W.

The individual MIRI filter catalogs are treated similarly except the separation thresholds increase more substantially with the increased growth rate of the PSF FWHMs. We adopt a 0.15\arcsec{} threshold between F770W and F1000W, 0.2\arcsec{} when adding F1500W and 0.25\arcsec{} for F2100W. These values were derived by inspecting a distribution of separation distances of matched sources between two catalogs and determining the point where mismatching interferes.

We combine the final NIRCam and MIRI catalogs with each other using a separation threshold of 0.3\arcsec{}. However the source density of the field is very high and dusty red objects only visible in the longest filters can easily be spuriously matched with young blue sources in the short NIRCam filters that happen to lie in a similar position along the line of sight. To mitigate this when combining the NIRCam and MIRI catalogs, we require that NIRCam sources matched to MIRI sources must have a F444W detection, which is our longest wavelength NIRCam filter. The young blue stars will not have an F444W detection in our data and therefore it acts as a good bridge between the near- and mid-IR.

We convert our final photometric catalog, which is generated in units of AB magnitudes by \textsc{starbugii}, into units of Vega magnitudes for easier comparison with previous work. For this conversion we employ the CRDS \texttt{jwst\_nircam\_abvegaoffset\_0001.asdf} and \texttt{jwst\_miri\_abvegaoffset\_0001.asdf} reference files. We apply a final foreground reddening correction to our NIRCam photometry. We adopt a value of $\mathrm{E(B-V)}=0.36$~\citep{Tantalo_2022} and $R_V=3.1$ to apply the extinction curve of \cite{Cardelli1989}. At mid-IR wavelengths, such foreground reddening is negligible, thus we apply no additional corrections to our MIRI photometry.


\begin{deluxetable*}{lc}
\label{tab:color_cuts}
\tabletypesize{\normalsize}
\tablewidth{0pt}
\tablecaption{
Color Selection Criteria Used to Identify Red Sources
}
\tablehead{\colhead{Color Selection}&\colhead{Number of Sources Selected} \\
}
\startdata
F770W$-$F1500W $\geq 0.6$ and $13.5 \leq \mathrm{F1500W} \leq 18$ & 56 \\
F1000W$-$F1500W $\geq 0.6$ and $13.5 \leq \mathrm{F1500W} \leq 18$ & 58 \\
F356W$-$F770W $\geq 1.0$ and $16.2 \leq \mathrm{F770W} \leq 20.5$ & 253 \\
F444W$-$F770W $\geq 1.0$ and $16.2 \leq \mathrm{F770W} \leq 20.5$ & 245 \\
$\mathrm{y} > -8.9\,\mathrm{x} + 33.7$ and $\mathrm{y} > -0.2\,\mathrm{x} + 18.0$ and $\mathrm{y} < -0.2\,\mathrm{x} + 23.5$ & 769$^{a}$ \\
$\mathrm{y} > -9.7\,\mathrm{x} + 34.9$ and $\mathrm{y} > -0.2\,\mathrm{x} + 17.7$ and $\mathrm{y} < -0.2\,\mathrm{x} + 23.6$ & 708$^{b}$ \\
\hline
Number of unique sources & 1,307 \\
\enddata
\tablecomments{$^{a}$Here y corresponds to the F444W magnitude, and x corresponds to the F115W$-$F444W color. $^{b}$Here y corresponds to the F356W magnitude, and x corresponds to the F115W$-$F356W color.}
\end{deluxetable*}

\section{Results} 
\label{sec:prelimresults}
\subsection{Images} 
\label{sec:images}
In the left panel of Figure \ref{fig:color_imgs}, we show a NIRCam three-color image of Spitzer~I, combining the F115W filter in blue, the F356W filter in green, and the F444W in red, while the right panel shows a MIRI three-color image combining F770W in blue, F1000W in green, and F2100W in red. At the shorter NIRCam wavelengths, Spitzer~I resembles a densely populated star field and little diffuse and extended emission is visible. We see only in the F356W filter (green) some very faint diffuse emission which appears to track the filaments seen in the MIRI images. The F356W filter is broad and overlaps with the F335M filter which traces the 3.3~\micron{} polycyclic aromatic hydrocarbon (PAH) feature \citep{sandstrom23}. The 3.3~\micron{} emission feature is thought to be the result of a C$-$H stretching vibration mode in small, neutral PAHs \citep{schutte93,vanDiedenhoven04} excited by UV photons. As Spitzer~I is a very active star-forming region \citep[][]{jones19,hirschauer20,kinson21}, the faint emission we see in the F356W filter may be associated with PAHs. A number of resolved background galaxies are also visible in the NIRCam image, most notably what appears to be a spiral galaxy close to the center.

In contrast to the abundance of stars we see in the NIRCam images, MIRI provides us with a different, complementary perspective of Spitzer I (Figure \ref{fig:color_imgs}, right panel). We observe widespread, diffuse emission in the F770W filter (blue). This band traces the 7.7~\micron{} PAH feature, which originates from the C$-$C stretching vibrational modes of larger, positively-charged molecules. Several features stand out in the F1000W filter (green), which traces warm dust from silicates, such as the spiral galaxy we observe in NIRCam close to the center of the image. Interestingly, we see in the top right corner of the MIRI image a point-like source with a pinwheel-like ``tail''. The pinwheel structure resembles that of the Wolf-Rayet (WR) star WR 104, which hosts an OB companion \citep[see Figure 1 of][]{tuthill99}. The collision of stellar winds produces dust \citep[see also][]{lau22} that is then radially swept by the WR wind into a stream that follows an Archimedian spiral. However, there are currently only four known WR stars in NGC~6822, none of which fall in our MIRI field-of-view \citep{massey1987}, and the large angular size of this object at the distance of NGC~6822 makes this unlikely. Another possibility is that this may be a supernova remnant (SNR) superimposed on a background galaxy, or it may simply be a background galaxy.

In the MIRI F2100W filter, we observe diffuse emission associated with some of the F770W emission; however, we also see a bubble structure toward the lower left of the image that is not visible in other filters in this data set. This corresponds to SNR Ho~12 which was first identified by \citet{hodge77}. \citet{kong04} performed a multi-wavelength analysis of this SNR, with X-ray, optical, and radio observations, and estimated the age of Ho~12 to be $1700-5800$ years. SNR Ho~12 will be investigated further using these MIRI observations in an upcoming work (Kavanagh et al.\ in prep.). 

\emph{JWST} has revealed the structure of the Spitzer~I star-forming region in exceptional detail, and in the following sections, we will investigate color-magnitude diagrams (CMDs) and SEDs of sources in this region extracted with \textsc{starbugii} \citep{nally22}.

\subsection{YSO Selection}
\subsubsection{Color Magnitude Diagrams}
\label{sec:CMDs}
\begin{figure*}
    \centering
    \includegraphics[width=0.41\textwidth]{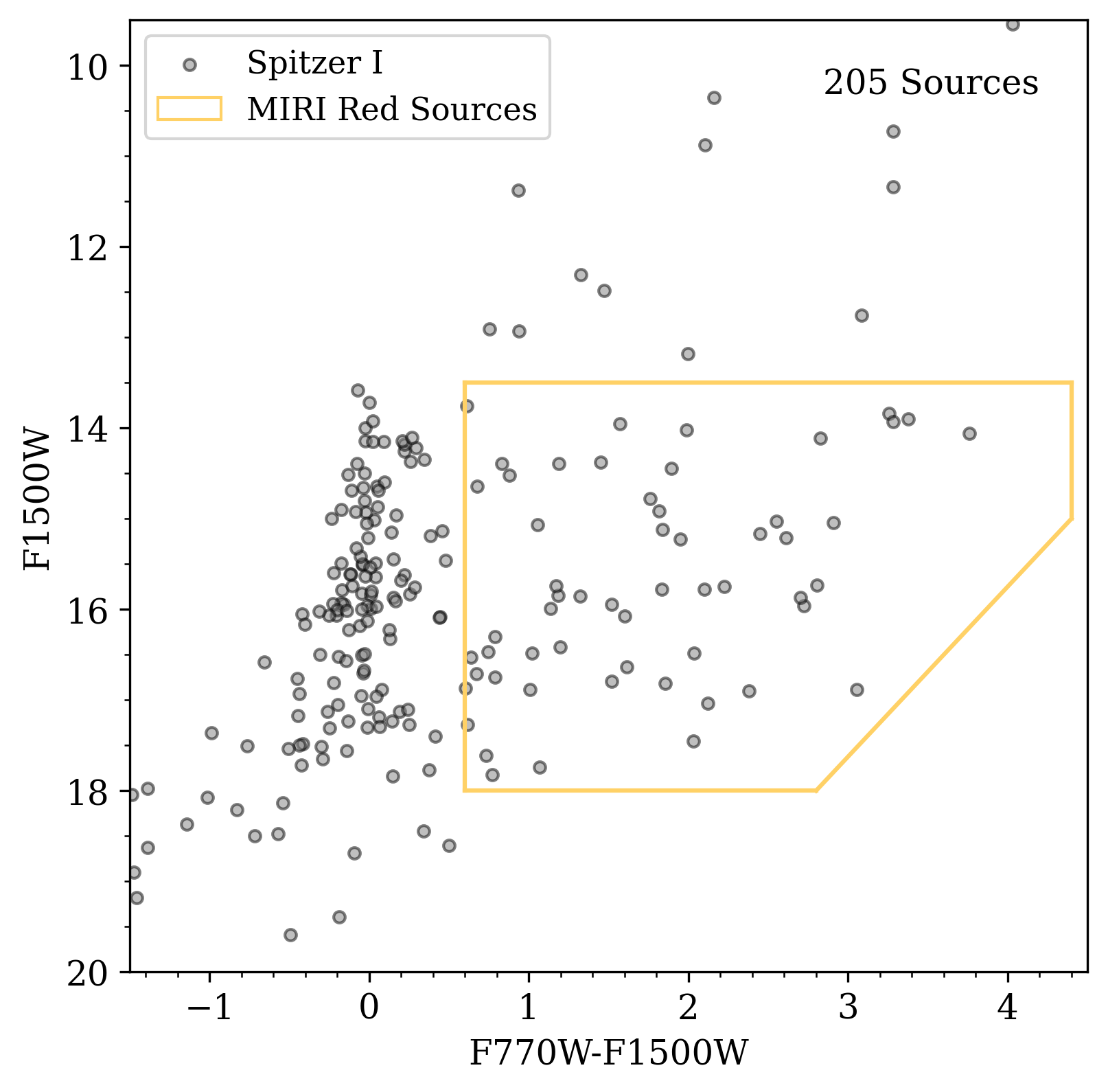}
    \includegraphics[width=0.41\textwidth]{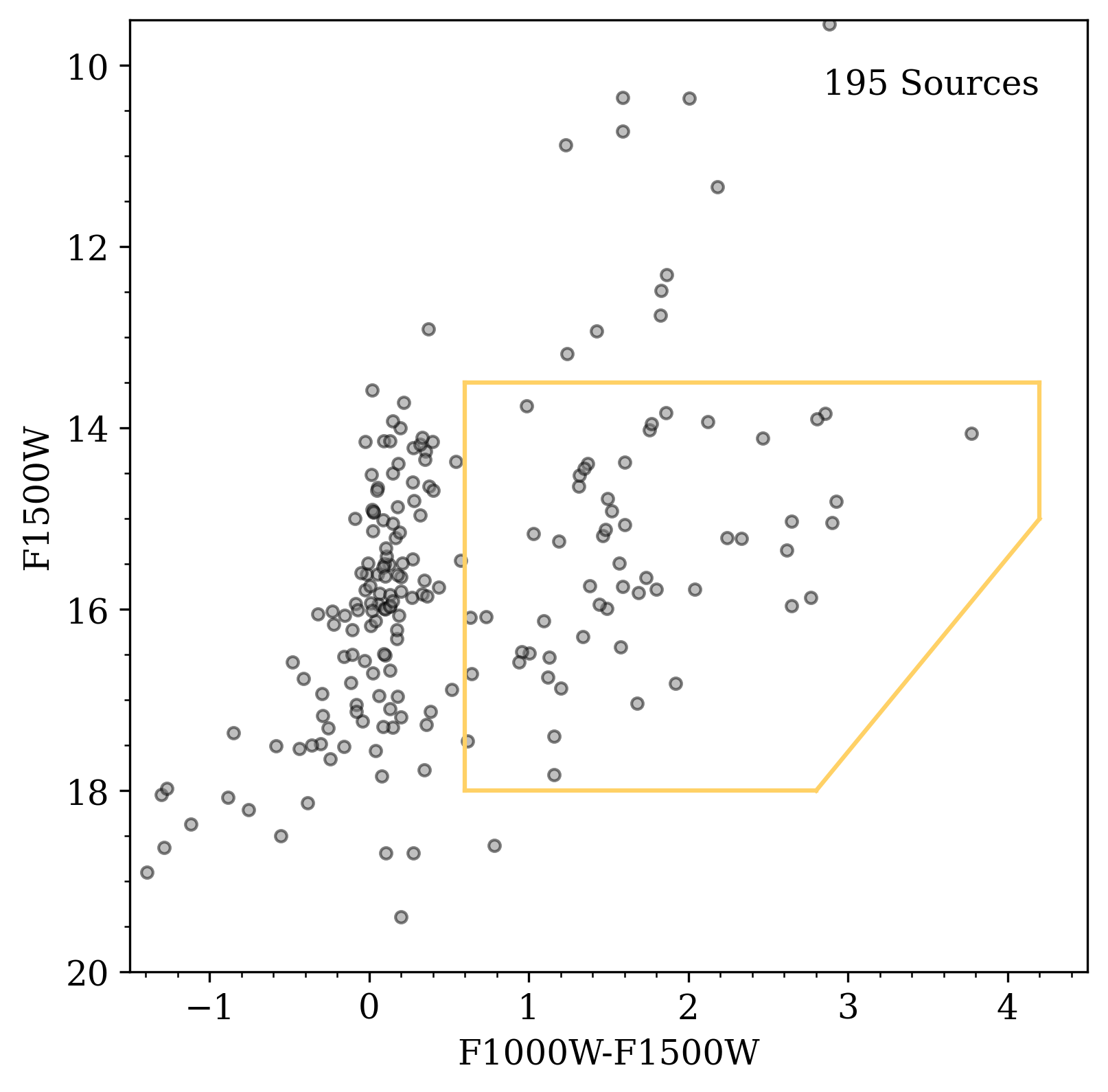}
    
    \includegraphics[width=0.41\textwidth]{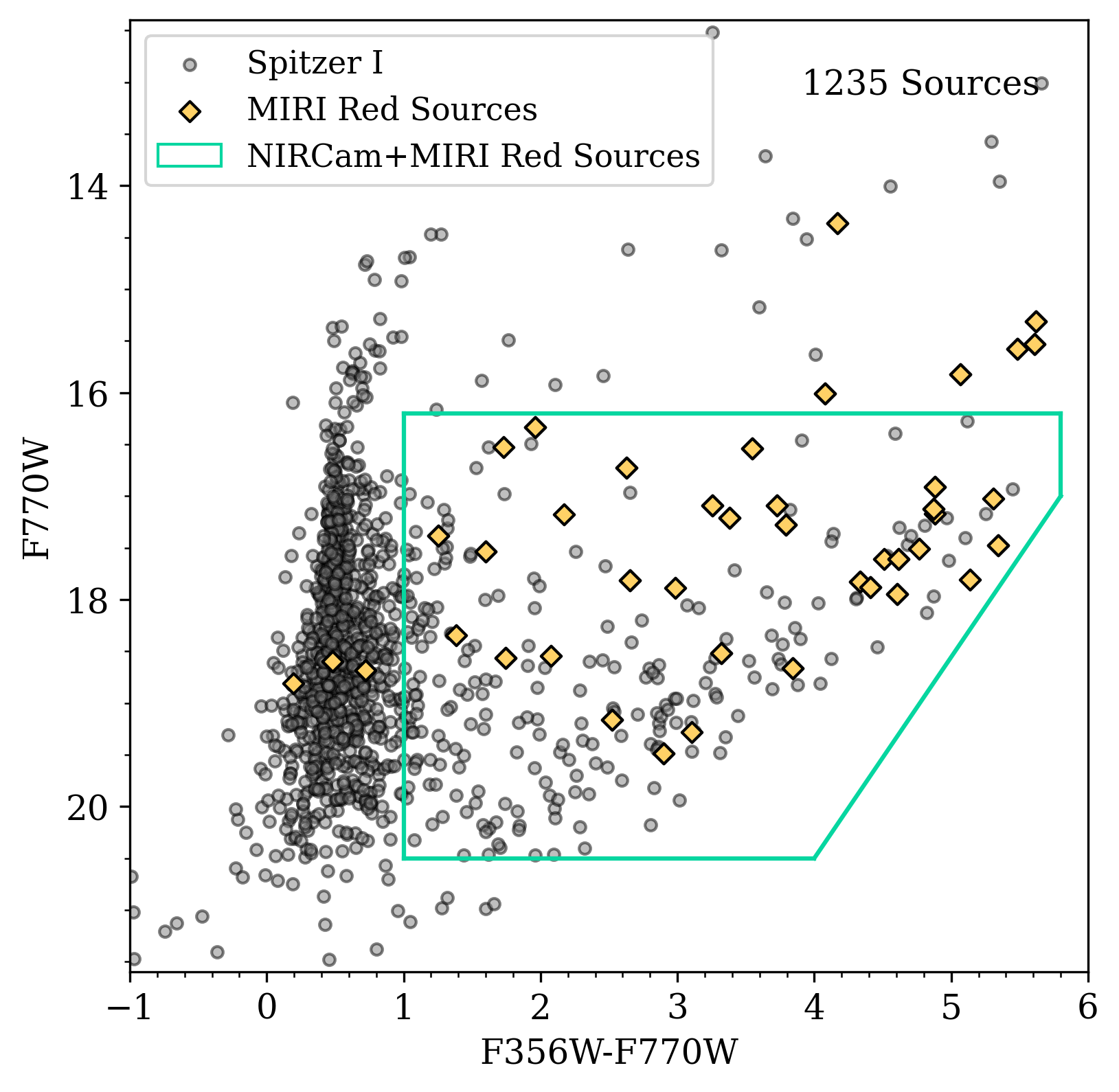}
    \includegraphics[width=0.41\textwidth]{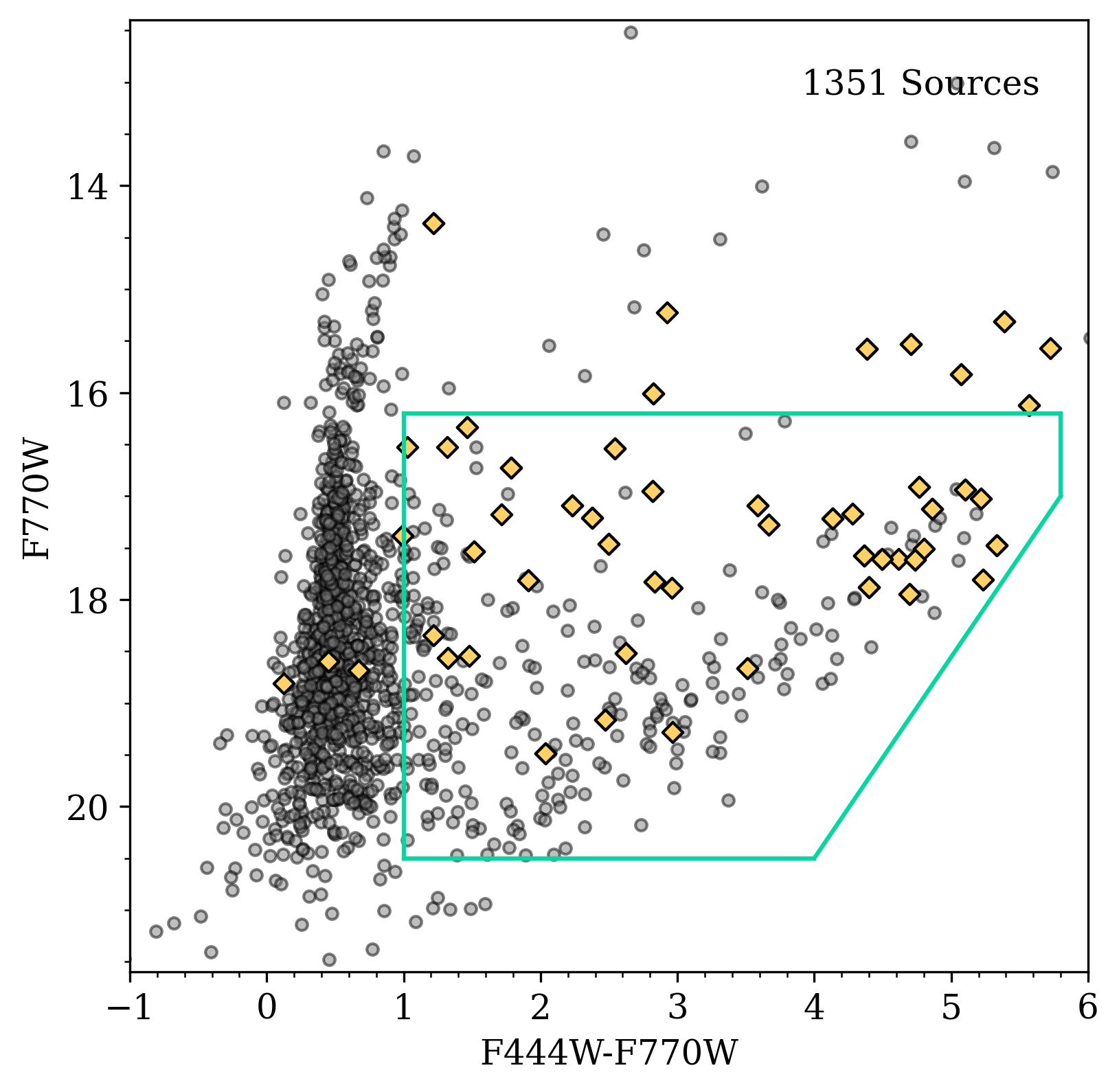}
    
    \includegraphics[width=0.41\textwidth]{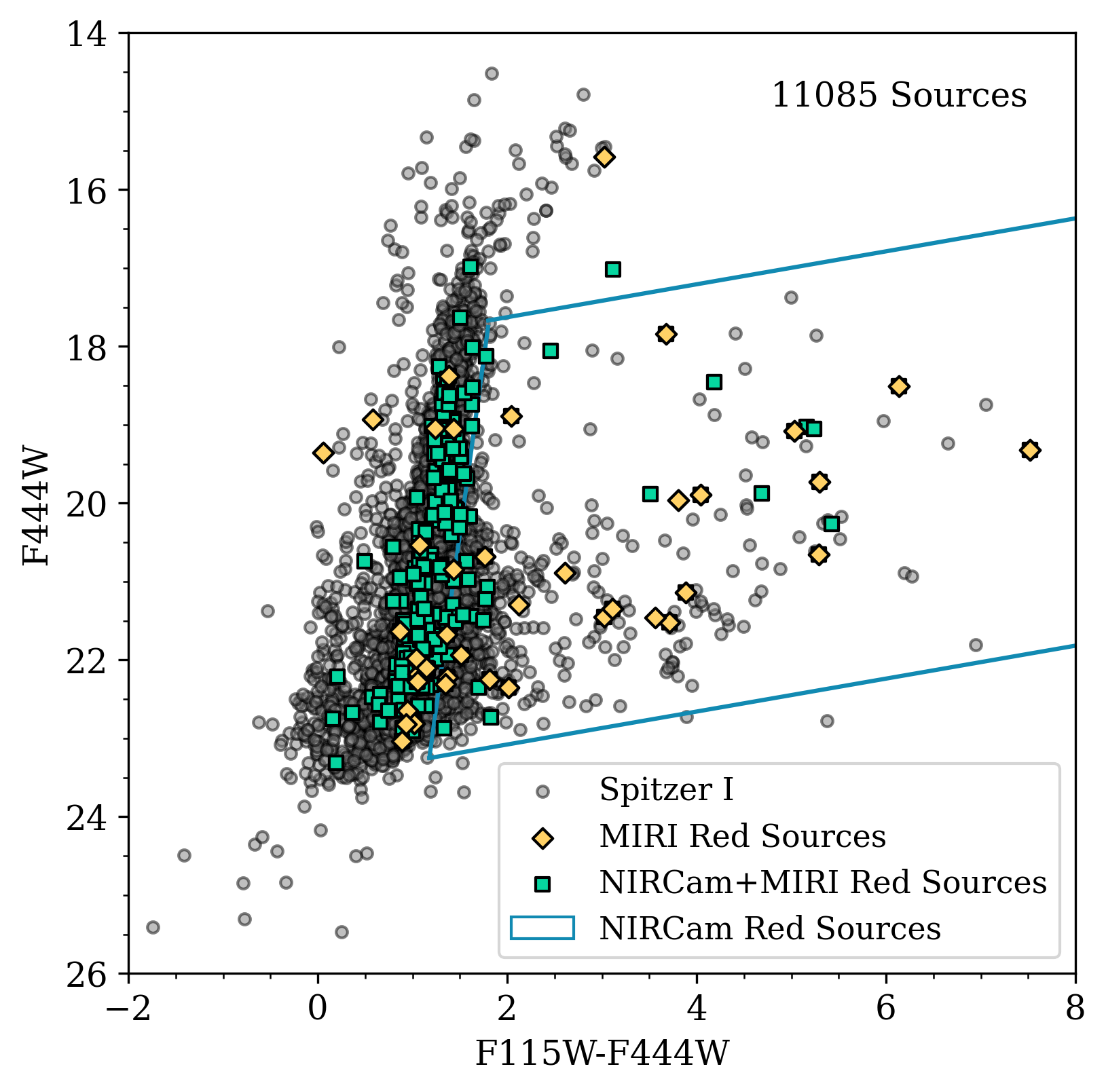}
    \includegraphics[width=0.41\textwidth]{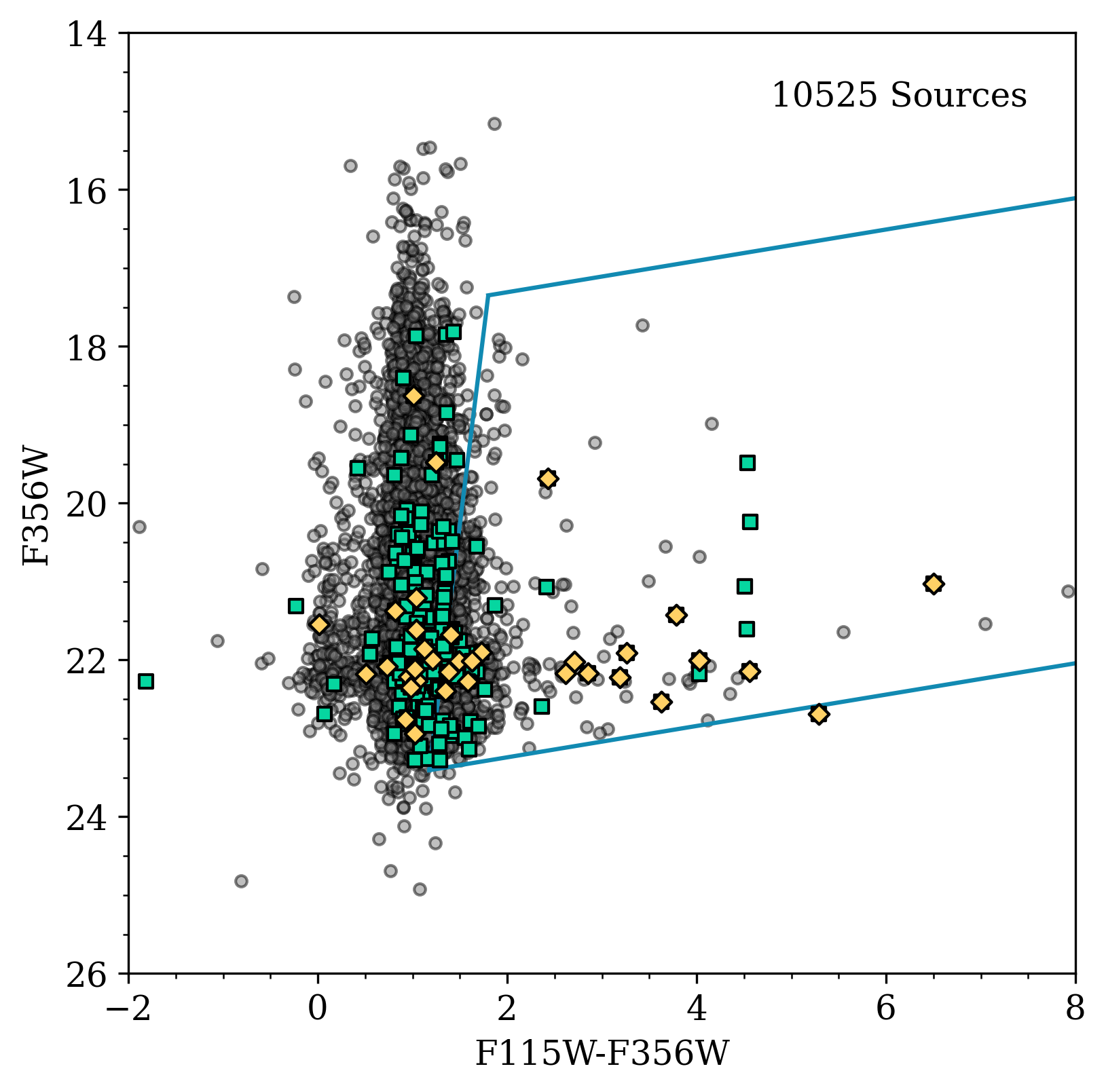}
    \caption{Color-magnitude diagrams (CMDs) of all sources within the Spitzer~I region (gray points). \textit{First Row:} F1500W versus F770W$-$F1500W (left) and F1500W versus F1000W$-$F1500W (right). The orange outline indicates ``red'' sources that are selected for further investigation. \textit{Middle Row:} F770W versus F356W$-$F770W (left) and F770W versus F444W$-$F770W (right). The teal outline indicates ``red'' sources that we select for further investigation while the yellow diamonds correspond to sources selected from the MIRI-only CMDs. \textit{Last Row:} F444W versus F115W$-$F444W (left) and F356W versus F115W$-$F356W (right). The blue outline indicates ``red'' sources that we select for further investigation while the yellow diamonds and teal squares correspond to sources selected from the previous four CMDs. The number of sources in each CMD is listed in the top right corner of every panel.}
    \label{fig:nircam_miri_cmds}
\end{figure*}

As Spitzer~I is likely the youngest and most active star-forming region in NGC~6822, we combine NIRCam and MIRI photometry to identify its YSO population, from the most embedded sources to those that are approaching the main sequence. In their earliest stages, protostars are associated with a dusty, infalling envelope and an accretion disk. As the protostar evolves, the disk and circumstellar envelope disperse, revealing the young star \citep{robitaille06}. Throughout this evolution, YSOs are seen at mid-IR wavelengths in their embedded phase, and emerge at shorter wavelengths when their disks and circumstellar dust dissipate or become optically thin.

Past studies have relied on \textit{Spitzer} IRAC and MIPS colors to select regions of color-magnitude space that are likely to be dominated by YSOs \citep[see, e.g.,][for the LMC, SMC, and NGC~6822 respectively]{Whitney2008,sewilo13,jones19}. In order to identify YSOs in Spitzer~I and characterize their SEDs across a range of evolutionary stages, we have constructed six CMDs from which we define provisional color cuts: We use F1500W versus F1000W-F1500W and F1500W versus F770W$-$F1500W to identify the most embedded sources, F770W versus F444W$-$F770W and F770W versus F356W$-$F770W to identify sources that are emerging from their circumstellar envelopes, and F444W versus F115W$-$F444W and F356W versus F115W$-$F356W to identify the most evolved YSOs with optically-thin or no disks. We present these CMDs in Figure \ref{fig:nircam_miri_cmds}, where the colored outlines indicate the color cuts we have defined through visual inspection. In each of these CMDs, we impose a limiting magnitude on the bright end to minimize contamination from dust-enshrouded asymptotic giant branch (AGB) stars \citep[see][for identification of AGB stars in NGC~6822]{nally2023}.

\begin{figure*}
    \includegraphics[width=\textwidth]{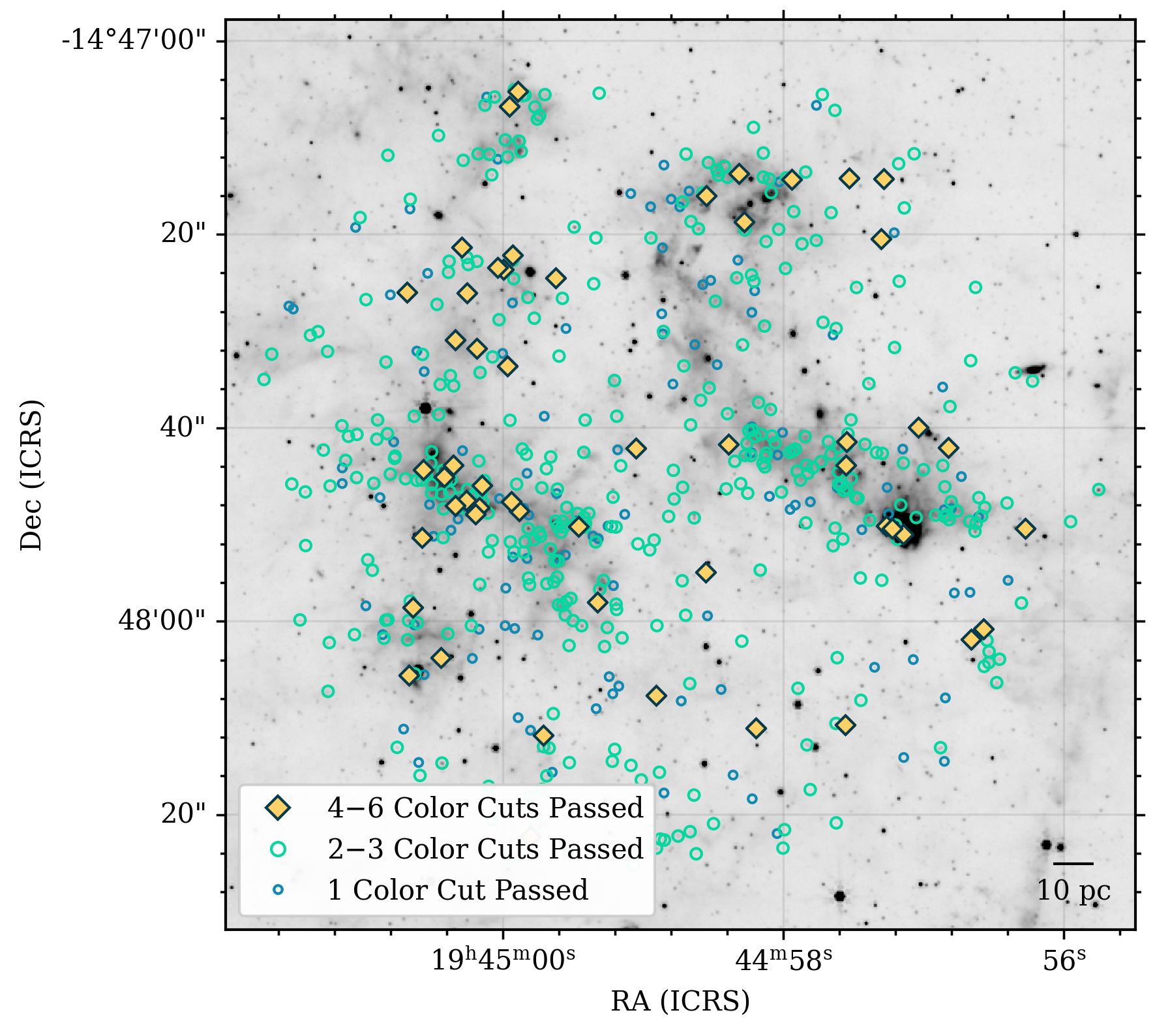}
    \caption{MIRI F770W image showing the spatial distribution of YSOs and candidate YSOs in the Spitzer~I region of NGC~6822. The image is centered at RA\,$=$\,19:44:58.7318 and Dec\,$=$\,$-$14:47:44.811 and north is up. The yellow diamonds correspond to sources that pass four to six color cuts, the teal circles are sources that pass two to three color cuts, and the smallest blue circles are sources that pass only one color cut. While YSOs and YSO candidates are distributed throughout the region, there are overdensities of sources along the dusty filaments.}
    \label{fig:spatial_dist}
\end{figure*}

We define the Spitzer~I star-forming region with a circular aperture of radius ${\sim} 46$\arcsec{} (109~pc) centered at RA\,$=$\,19:44:59.0160 and Dec\,$=$\,$-$14:47:40.920 \citep[this is the same definition of Spitzer~I as in][]{jones19}. There are 81,913 sources in this region \citep[][]{nally2023} and after applying the color cuts we define in Table \ref{tab:color_cuts}, 1,307 of these are classified as ``red'' sources. We show the spatial distribution of the subset of sources that are classified as YSOs or YSO candidates over the MIRI F770W image in Figure \ref{fig:spatial_dist} (see Section \ref{subsec:sed_fit}). Here, the yellow diamonds represent objects that show an IR excess in four to six colors, open teal circles are objects that show an IR excess in two to three colors, and open blue circles are objects that possess an IR excess in only one color. We observe red sources distributed throughout the Spitzer~I region; however, there are clear overdensities of sources associated with the two prominent filaments that characterize the morphology of Spitzer~I at the longer MIRI wavelengths.

\begin{deluxetable*}{lcccccccc}
\label{tab:props}
\tabletypesize{\normalsize}
\tablewidth{0pt}
\tablecaption{Physical properties of \emph{JWST}-Identified YSOs in Spitzer~I}
\tablehead{
    \colhead{Source}        &
    \colhead{RA}            &
    \colhead{Dec.}           &
    \colhead{Radius, $R_{*}$}       &
    \colhead{$R_{*}$ Error}       &
    \colhead{Temperature, $T_{\mathrm{eff}}$} &
    \colhead{$T_{\mathrm{eff}}$ Error} &
    \colhead{Luminosity, $L_{*}$}       &
    \colhead{Mass, $M_{*}$}       \\
    \colhead{}              &
    \colhead{(J2000)}       &
    \colhead{(J2000)}        &
    \colhead{[R$_{\odot}$]} &
    \colhead{[R$_{\odot}$]} &
    \colhead{[K]}           &
    \colhead{[K]}           &
    \colhead{[L$_{\odot}$]} &
    \colhead{[M$_{\odot}$]}\\
}
\startdata
CN233 & 296.235730 & -14.800212 & 8.13 & 0.99 & 4701 & 2378 & 29 & 2.62 \\
CN282 & 296.236087 & -14.800518 & 2.73 & 0.30 & 10170 & 2260 & 72 & 3.39 \\
CN1878 & 296.239839 & -14.802982 & 69.70 & 13.80 & 2565 & 271 & 190 & 4.47 \\
CN5999 & 296.251849 & -14.801038 & 9.80 & 2.68 & 7895 & 1294 & 336 & 5.27 \\
CN7649 & 296.251480 & -14.795518 & 8.10 & 2.33 & 9843 & 1480 & 555 & 6.08 \\
CN7761 & 296.237657 & -14.794440 & 42.54 & 1.21 & 3348 & 1866 & 205 & 4.57 \\
CN8224 & 296.239725 & -14.787267 & 21.05 & 2.93 & 6296 & 583.5 & 627 & 6.30 \\
CN8344 & 296.239821 & -14.795527 & 69.05 & 9.65 & 4478 & 526 & 1727 & 8.41 \\
CN9099 & 296.242984 & -14.787147 & 54.75 & 5.21 & 8787 & 220 & 16100 & 15.92 \\
CN10964 & 296.249995 & -14.789897 & 83.98 & 1.04 & 2518 & 1393.5 & 255 & 4.87 \\
CN11125 & 296.251424 & -14.791925 & 22.63 & 0.76 & 7645 & 301 & 1576 & 8.20 \\
CN11282 & 296.250783 & -14.792168 & 16.94 & 2.37 & 4071 & 910 & 71 & 3.38 \\
\enddata
\tablecomments{The 12 sources listed here correspond to the sources shown in Figure \ref{fig:seds}. The full table is available online. Column 1: Arbitrary catalog number. Columns $2-3$: Right ascension and declination of the source in degrees. Column 4: Best-fit radius of the YSO candidate. Column 5: Radius median absolute deviation of all fits that are within a reduced $\chi^{2}$ value of $<3$ from the best fit. Column 6: Best-fit temperature. Column 7: Temperature median absolute deviation of all fits that are within a reduced $\chi^{2}$ value of $<3$ from the best fit. Column 8: Luminosity of the source calculated from the best-bit radius and temperature: $\mathrm{L} = 4 \pi \mathrm{R}^{2} \sigma \mathrm{T}^{4}$. Column 9: Mass of the YSO candidate obtained from the luminosity -- $\mathrm{L} \propto \mathrm{M}^{3.5}$.}
\end{deluxetable*}

\subsubsection{YSO Spectral Energy Distribution Fitting} \label{subsec:sed_fit}
It is important to note that selecting red sources via CMD color cuts is prone to contamination from evolved stars and background galaxies. The MIRI filters in this case are crucial to mitigate the contamination from evolved stars \citep{Jones2017}. However, as the sensitivity limit at longer wavelengths decreases, the source counts at subsequently longer wavelength MIRI filters decreases as well. This limits the number of sources where we can use long baseline MIRI colors to minimize the contamination from evolved stars. To clean our red source catalog and identify the YSOs and YSO candidates, we produce and visually inspect the SEDs for each of the 1,307 red objects. In addition, we visually inspect zoom-in cutout regions centered on the RA and Dec.\ coordinates of each source. In doing this, we identify extended sources that resemble background galaxies and remove them from our YSO candidate list. Similarly, we identify and remove sources whose SEDs show decreasing flux with increasing wavelength, consistent with evolved stars. Finally, we additionally fit every SED with the ``spubhmi'' YSO model SEDs from \citet{Robitaille2017} with the SED Fitter tool \citep{Robitaille2007}, convolved to the JWST filter set by \citet{Richardson2024}\footnote{We used the version 1.1 model set \citep{richardson_2023_8356472}}. 

After fitting the \citet{Robitaille2017} YSO model SEDs to all 1,307 objects in our red source catalog, we use the following criteria to separate YSOs and YSO candidates from contaminating evolved stars and background galaxies: (1) a source that appears point-like in all images and is well-fit \citep[reduced $\chi^{2} \leq 10$; as was done in][]{nayak23} by a model YSO SED is classified as a YSO; (2) point-like sources with poorer SED fits \citep[$10 < \mathrm{reduced}\, \chi^{2} \leq 15$; similar to][]{jones19} are classified as YSO candidates; (3) point-like sources with SEDs that drop off at longer wavelengths and are poorly fit by the YSO model SEDs (reduced $\chi^{2} > 15$) are classified as evolved stars; (4) extended sources are classified as background galaxies; and (5) all other sources that do not fit into the previous categories are left unclassified; these tend to be sources with too few photometric parameters for establishing a robust classification. 

\begin{figure*}
    \centering
    \includegraphics[width=0.32\textwidth]{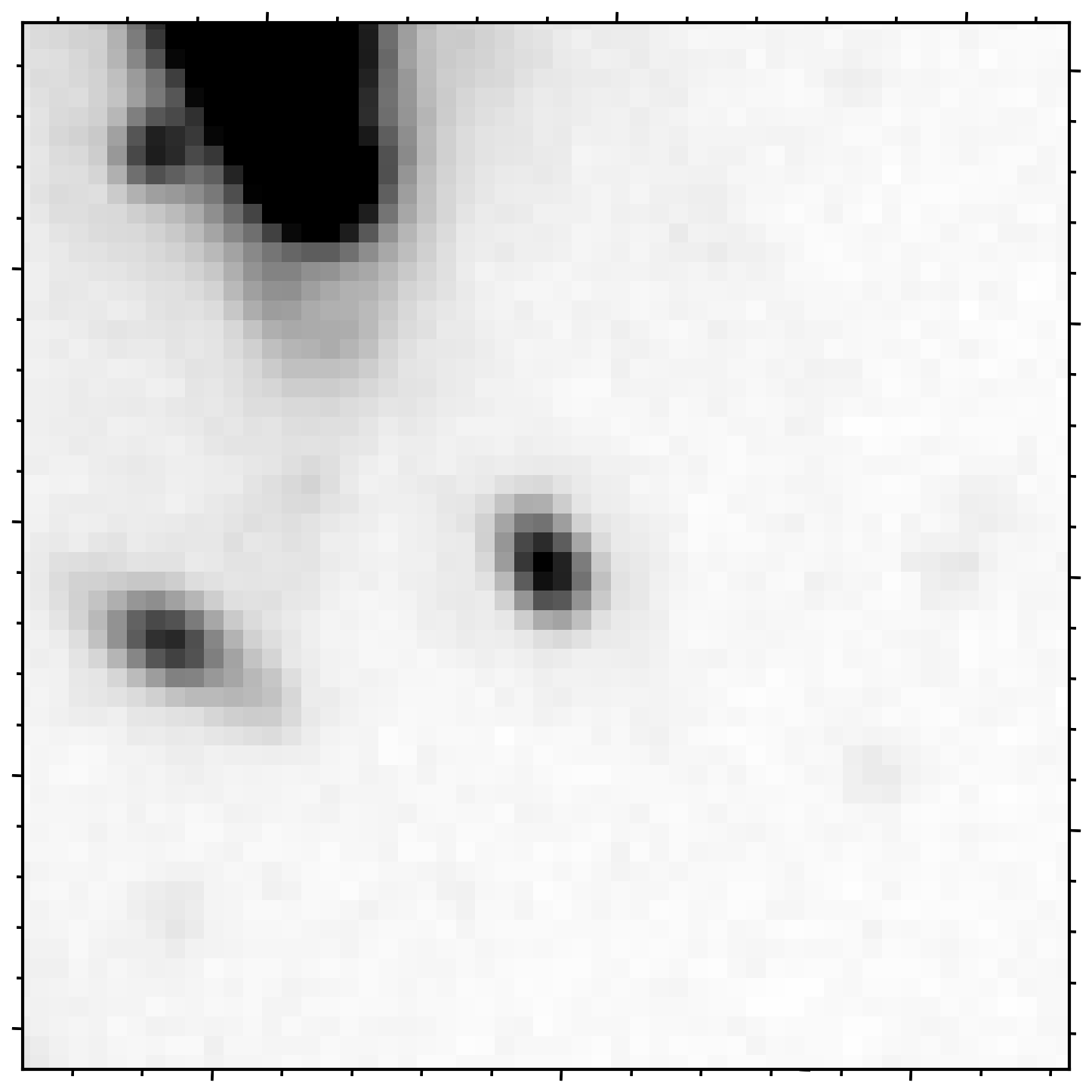}
    \includegraphics[width=0.32\textwidth]{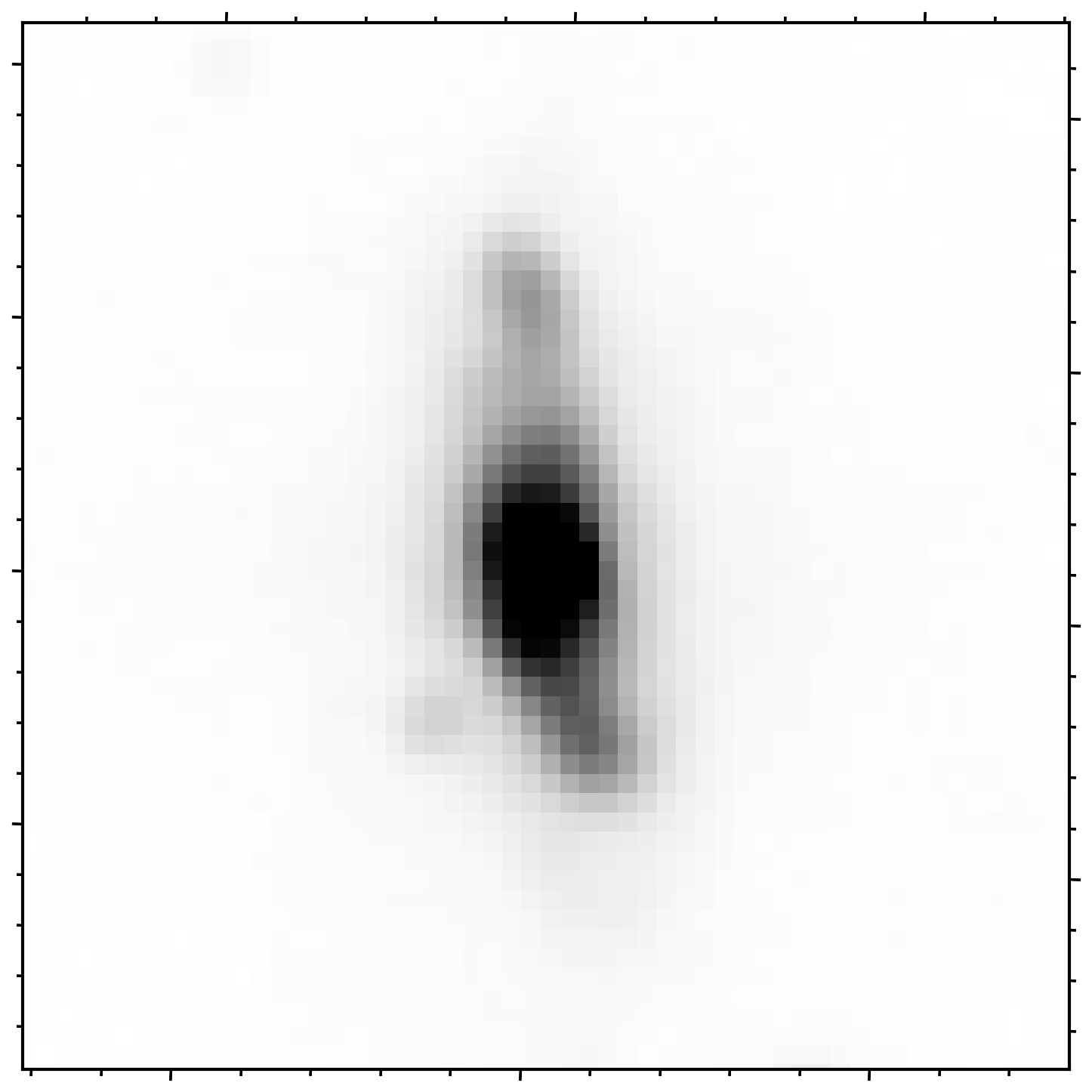}
    \includegraphics[width=0.32\textwidth]{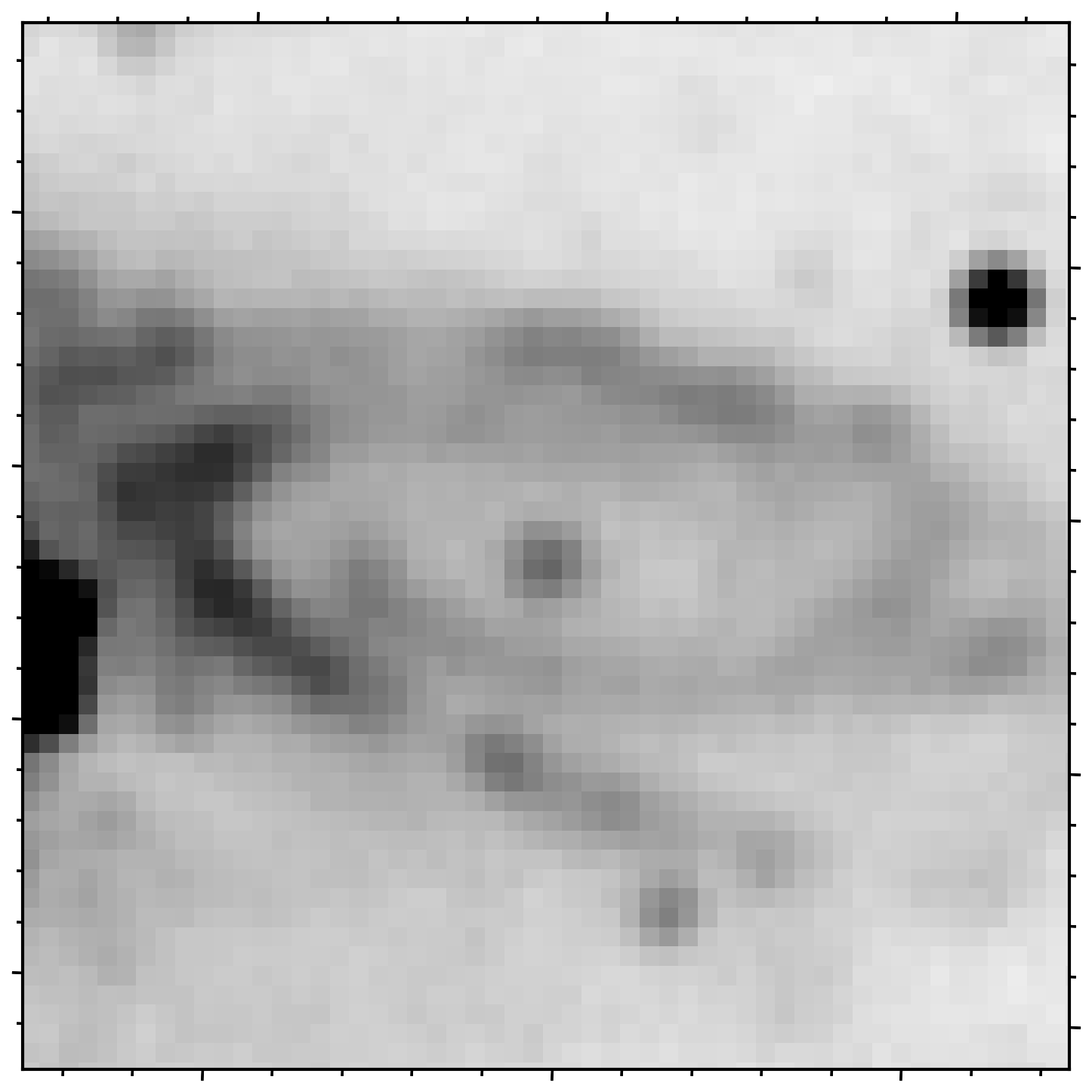}
    \caption{MIRI F1000W $6\times6$~arcsec$^{2}$ images centered on the three sources which passed all six color cuts, but are discarded from our YSO candidate list as background galaxies. The object in the left panel has an extended disk in all four NIRCam filters. The object in the right panel was classified as a YSO in \citet{jones19} and \citet{kinson21}, but with the resolution of \emph{JWST} we clearly identify spiral structure indicative of a background galaxy.}
    \label{fig:bkg_gal}
\end{figure*}

\begin{figure*}
    \centering
    \includegraphics[width=0.3\textwidth]{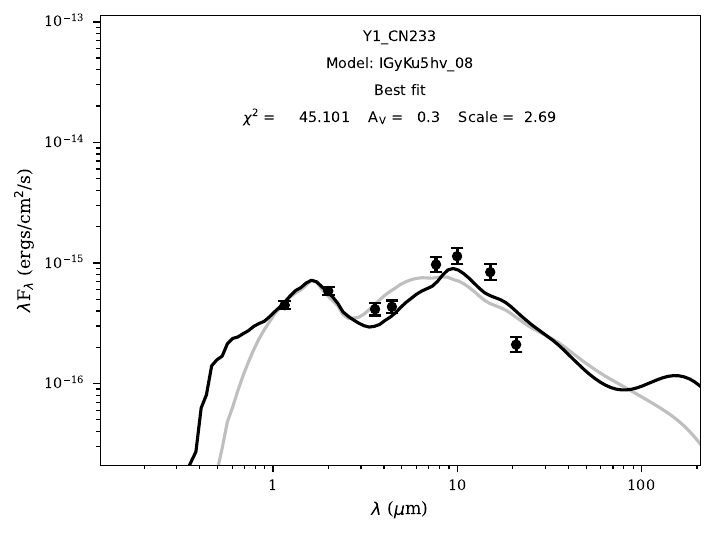}
    \includegraphics[width=0.3\textwidth]{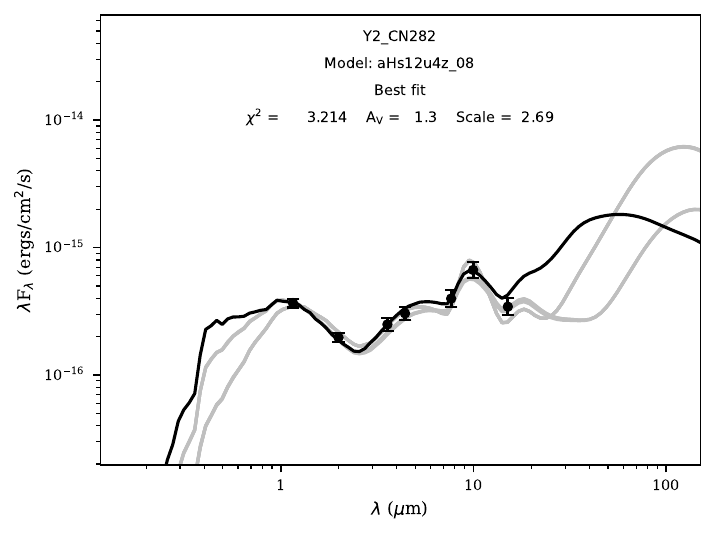}
    \includegraphics[width=0.3\textwidth]{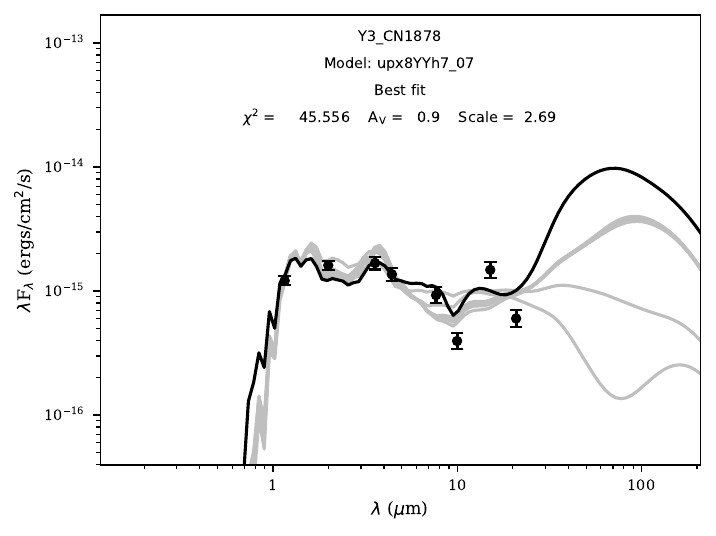}
    
    \includegraphics[width=0.3\textwidth]{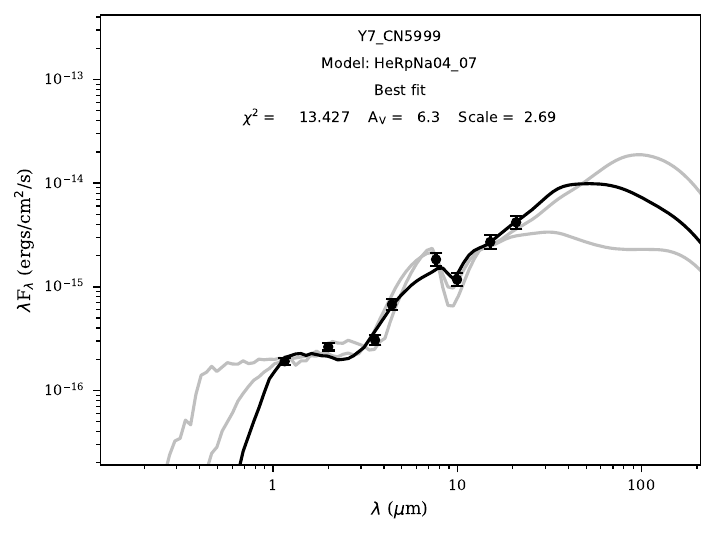}
    \includegraphics[width=0.3\textwidth]{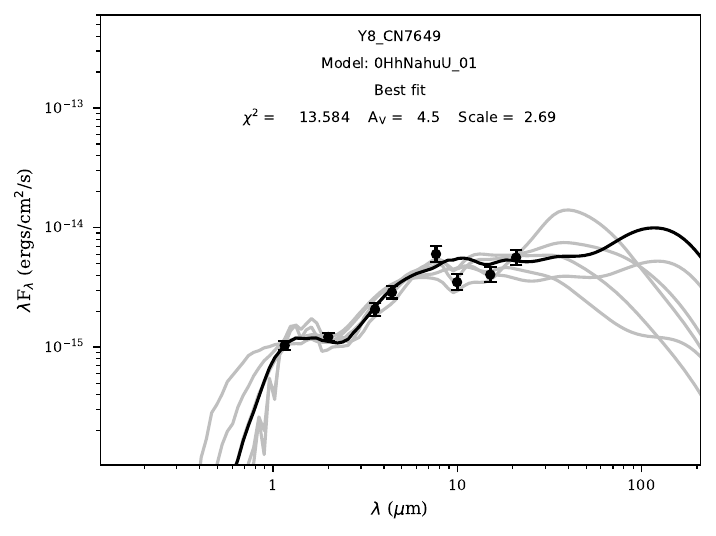}
    \includegraphics[width=0.3\textwidth]{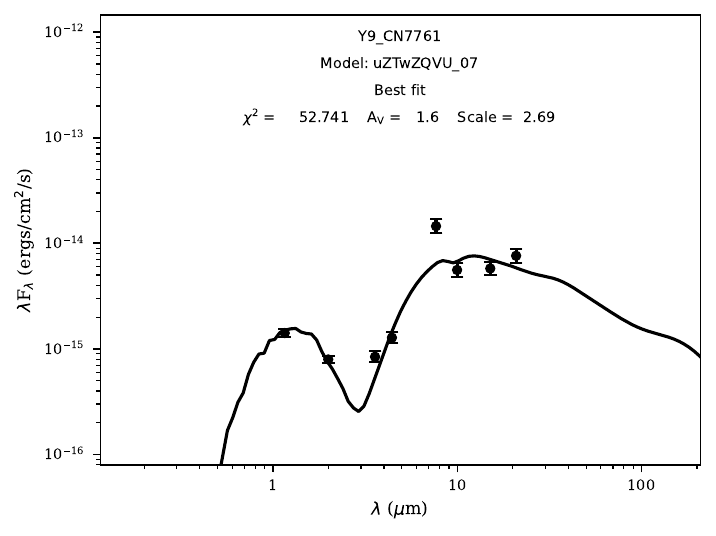}

    \includegraphics[width=0.3\textwidth]{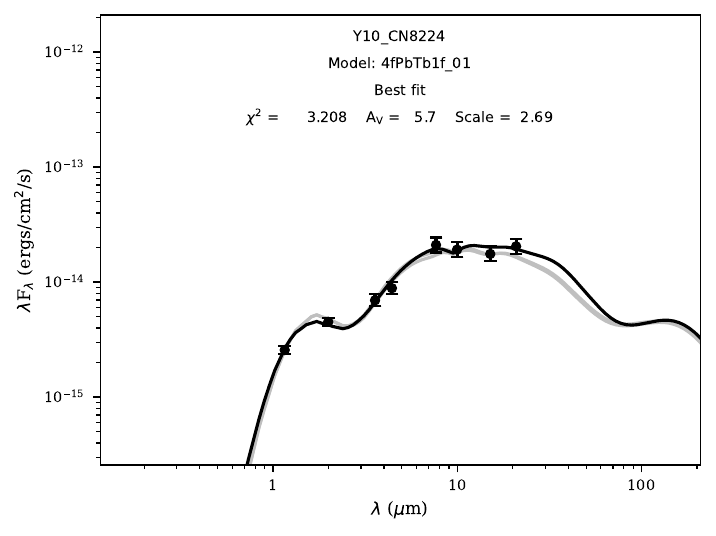}
    \includegraphics[width=0.3\textwidth]{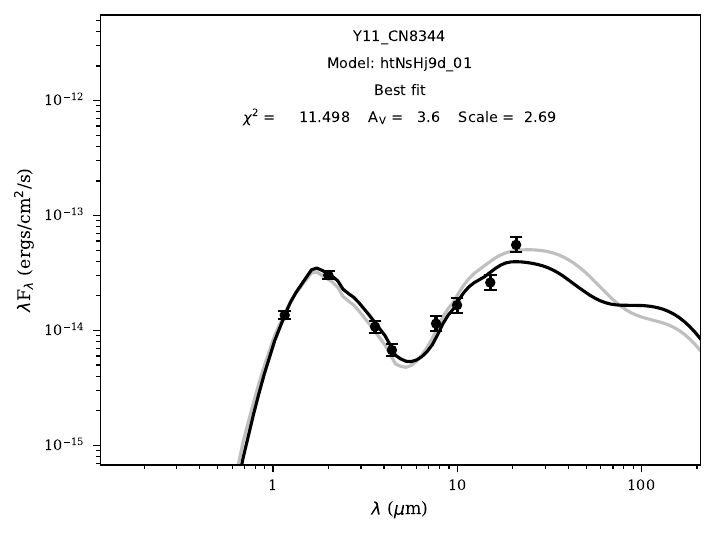}
    \includegraphics[width=0.3\textwidth]{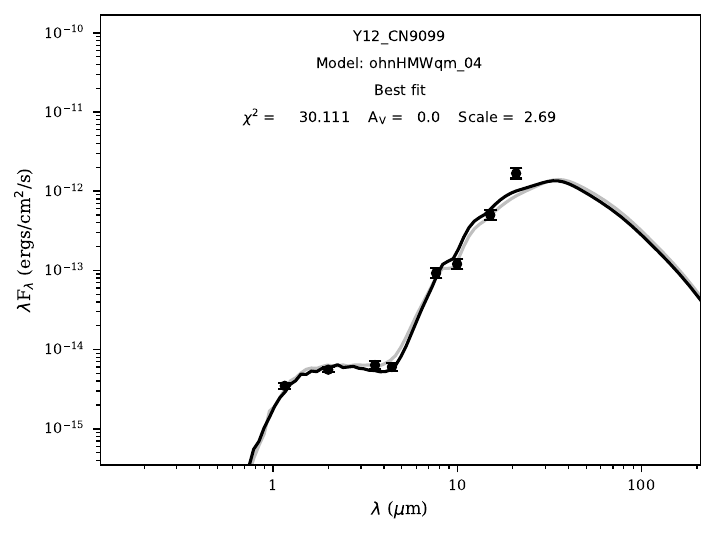}

    \includegraphics[width=0.3\textwidth]{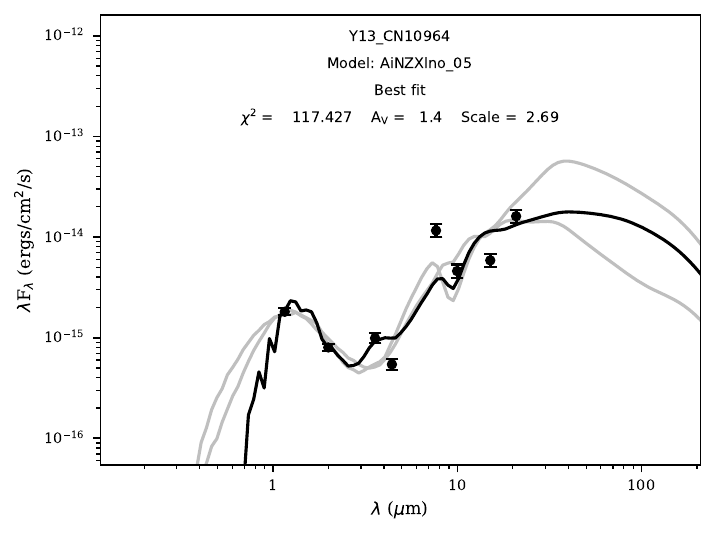}
    \includegraphics[width=0.3\textwidth]{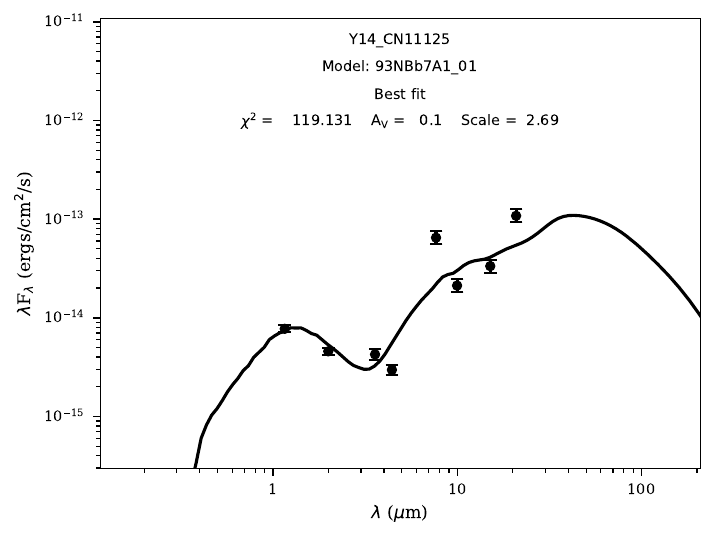}
    \includegraphics[width=0.3\textwidth]{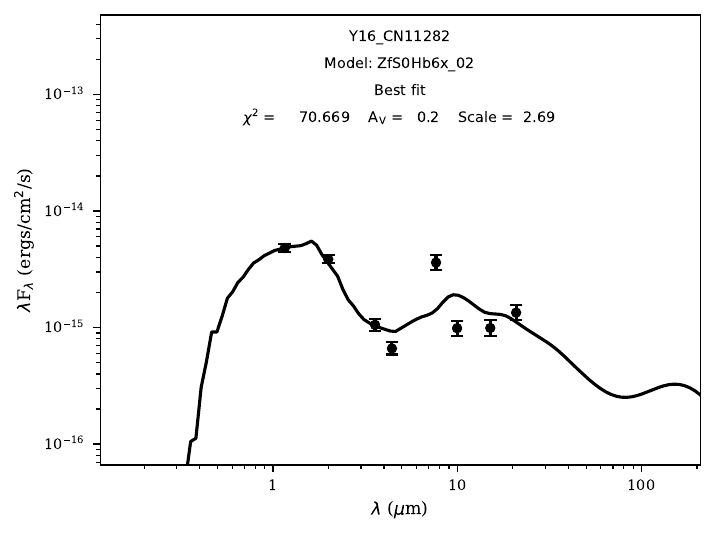}

    \caption{SED fits for the 12 objects selected as red sources in all six of our CMDs, where the black line is the best fit SED and the grey lines are fits with $\Delta\chi^{2} < 0.5$ with respect to the best fit. Each panel lists the best-fit model and the $\chi^{2}$ value of that best fit (not the reduced $\chi^{2}$). Visual inspection of these sources reveals that they are not background galaxies and their photometry is well fit by YSO models.}
    \label{fig:seds}
\end{figure*}

Of the 1,307 red sources we select from the CMDs, 16 objects satisfy all six color criteria that we describe in Table \ref{tab:color_cuts}. These 16 sources have photometry in all NIRCam and MIRI filters in this program, except for one which has no matched detections in the longest wavelength filter, F2100W. Visual inspection of the images of these 16 sources in all available filters reveals that four are background galaxies; we show examples of contaminating background galaxies in Figure \ref{fig:bkg_gal}. Of the remaining 12 sources, 10 are classified as YSOs and two are classified as YSO candidates; we show these in Figure \ref{fig:seds}. There are seven objects within our red source list that satisfy five out of the six color criteria we impose in Table \ref{tab:color_cuts}; two are identified as background galaxies, two are identified as YSOs, and three are identified as YSO candidates. There are 44 sources that pass four of our six color criteria: We identify two as evolved stars, 32 as YSOs, four as YSO candidates, and six sources as background galaxies. There are 13 sources that pass three of our color cuts; 11 are classified as YSOs, and two are classified as YSO candidates. There are 554 sources that pass two color cuts; 167 are classified as evolved stars, 77 are classified as YSOs, 279 are classified as YSO candidates, four are classified as background galaxies, and the remaining sources are unclassified. Finally, there are 683 sources that pass one color cut; 178 of these are classified as evolved stars, eight are classified as YSOs, 110 are classified as YSO candidates, 10 are classified as background galaxies, and the remaining sources are unclassified. We thus identify 140 YSOs and 400 YSO candidates; we list the properties of the 12 YSOs that satisfy all six color criteria in Table \ref{tab:props}, while a machine-readable version of the entire catalog will be available online. In total, we identify 26 background galaxies, which are all well-fit by YSO model SEDs (reduced $\chi^{2} < 10$); therefore, ${\sim 15}$\% of our ``YSO sample'' (red sources that are well-fit by YSO SEDs) is contaminated by background galaxies, and the high-resolution imaging provided by JWST allows us to eliminate these from our final YSO catalog.

\section{Discussion} \label{sec:discussion}
\subsection{Comparison of Spitzer-Identified YSOs to JWST}
Recently, \citet{jones19} studied the population of massive YSOs in NGC~6822 using archival \emph{JHK} observations with the United Kingdom Infrared Telescope \citep[UKIRT;][]{Sibbons2012}, and IRAC and MIPS observations from the \textit{Spitzer Space Telescope}. With CMD cuts and SED fitting, they identified 105 high-confidence and 88 medium-confidence YSOs associated with seven high-mass star-forming regions. These include Hubble I/III, IV, V, and X, but also the discovery of three new embedded high-mass star-forming regions: Spitzer I, II, and III. Of all star-forming regions in NGC~6822, Spitzer~I hosts the greatest number of color-selected YSOs \citep[90; see Table 9 in][]{jones19}. It is bright at IR wavelengths of 8 and 24~\micron{}, but faint in \HA{} and UV emission, suggesting that Spitzer~I may be the youngest and most actively star-forming region in NGC~6822. Furthermore, the high IR flux which traces embedded star formation, compared to \HA{} or UV which trace star formation over the lifetimes of O- and B-type stars, suggests that its star-formation activity may not have yet reached its peak \citep[as is seen in the proto-SSC H72.97–69.39;][]{ochsendorf17}.

A subsequent study by \citet{hirschauer20}, which investigates new methods for establishing color cuts through implementation of kernel density estimate techniques, employed the same archival \emph{JHK} UKIRT and \textit{Spitzer} IRAC and MIPS data to confirm the large presence of YSOs in Spitzer I.
Furthermore, they postulated that this region may be a proto-SSC ($M_{*} \gtrsim 10^{5}$~M$_{\odot}$). SSCs represent an extreme mode of star formation where the stellar surface densities exceed those of \HII{} regions and OB associations \citep{nayak19} by orders of magnitude and the star formation efficiencies are high \citep[see, e.g.,][]{turner15,oey17}. Within extragalactic members of the Local Group, the R136 cluster in the LMC's 30 Doradus region is the best-studied SSC \citep[see, e.g.,][]{Chevance2020,Wong2022}. SSCs are also observed in the central starburst of NGC~253 \citep{leroy18,levy21}, M82 \citep{smith06}, and NGC~4945 \citep{emig20}. Recent work has also revealed a proto-SSC (H72.97-69.39) in the largest of the three molecular clouds in the N79 region of the LMC \citep{ochsendorf17,nayak19}.

Using a supervised machine learning approach, \citet{kinson21} trained a probabilistic random forest classifier to re-investigate the YSO population within the star-forming bar of NGC~6822. Applying this method to the $J-H$, $H-K$, $J-K$ near-IR colors, $K-$band magnitudes from UKIRT, and far-IR surface brightness measurements at 70 and 160~\micron{} from \textit{Spitzer} and \textit{Herschel}, the authors identified 199 new YSOs, and confirmed the YSO nature of 125 out of the 277 literature YSO candidates from \citet{jones19} and \citet{hirschauer20}. They also definitively classified 82 of the 277 literature candidates as non-YSOs. Thus, their final catalog consists of 324 sources across the seven Hubble and Spitzer regions (Hubble I/III, IV, V, and X, and Spitzer I, II, and III), and four new sites of star formation that they identified through their work (BHD 9/10, 18, 27, and HIV-N). Of these 324 sources, \citet{kinson21} identified 90 YSOs and 49 YSO candidates within Spitzer~I (see their Table 4).

As \citet{jones19} discuss (see their Section 2.3), the limited resolution of \textit{Spitzer} can confuse color selections, and sources that are identified as YSO candidates from these data likely correspond to multiple objects rather than individual ones. To illustrate, the FWHM of \textit{Spitzer} at 3.6 and 24~\micron{} is 1.7\arcsec{} and 6\arcsec{} respectively; these correspond to 4 and 14.2~pc at the distance of NGC~6822 (490~kpc). In contrast, the angular resolution of our NIRCam observations ranges $0.04-0.145$\arcsec{} (${\sim} 0.1-0.3$~pc), while MIRI ranges $0.269-0.674$\arcsec{} (${\sim} 0.6-1.6$~pc). The UKIRT WFCAM $K-$band median image quality of 0.7\arcsec{} \citep[][]{Casali2007}, which corresponds to ${\sim} 1.7$~pc, is most comparable to our longest wavelength MIRI data (F2100W), where the effects of blending already become apparent when comparing to NIRCam. Therefore, with the improved resolution of \emph{JWST}, we are now able to revisit previously-identified YSOs and YSO candidates and investigate to what extent these limitations affected earlier studies. 

\begin{figure}
    \centering
    \includegraphics[width=\columnwidth]{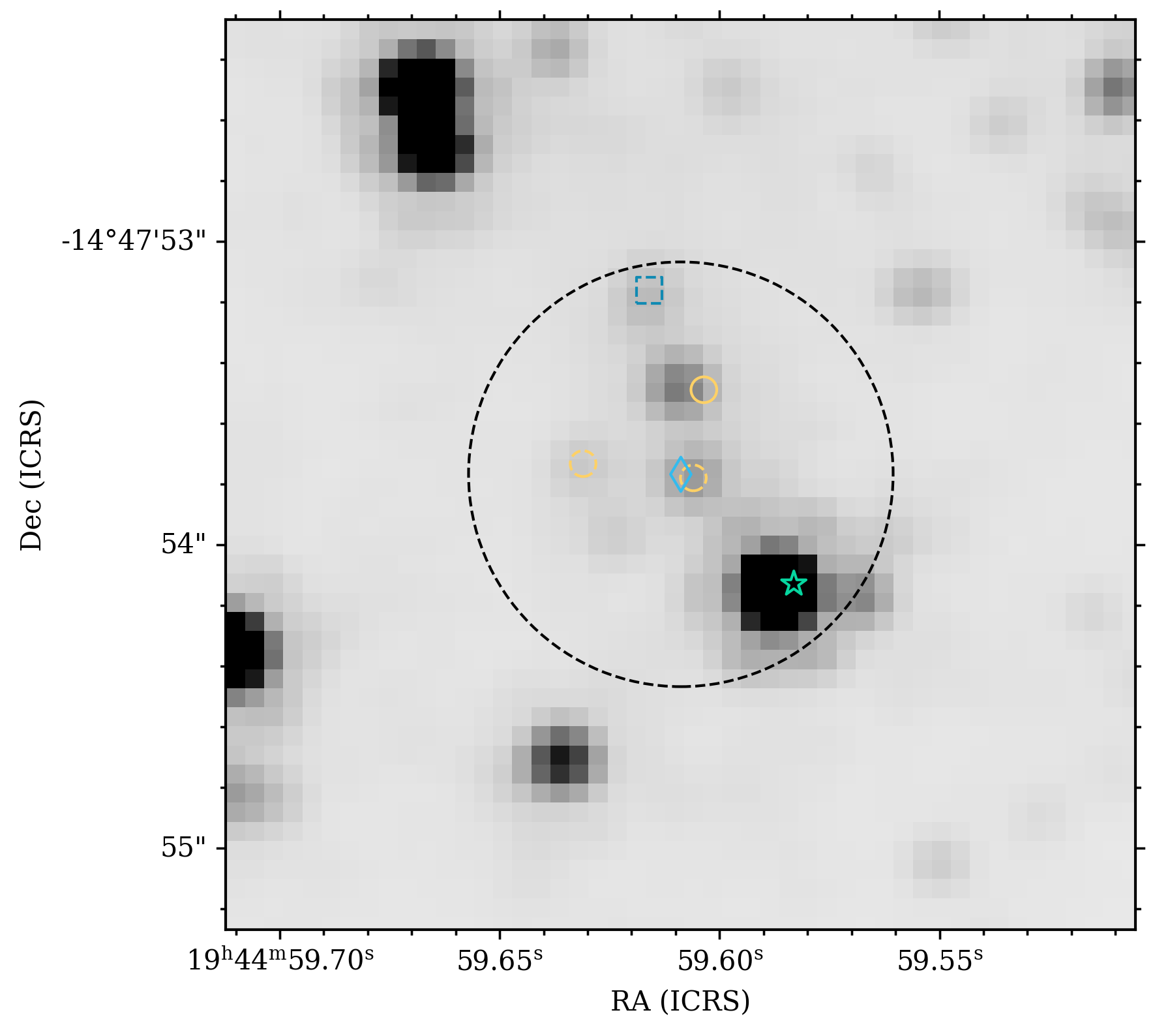}
    \caption{NIRCam F444W image showing a $3\times3$~arcsec$^{2}$ zoom-in on one YSO from the \citet{kinson21} catalog (cyan diamond symbol). There are \textbf{five} objects in our red source catalog that lie within 1\arcsec{} of it: We classify \textbf{one} of these as an evolved star (teal star symbol), three as YSOs and YSO candidates (yellow  circle symbols), and on is unclassified (blue square). At the resolution of \textit{Spitzer} IRAC and MIPS, these sources would not have been separately resolved within the larger PSFs.}
    \label{fig:mult_match}
\end{figure}

We first compare our catalog of red sources to the work of \citet{kinson21}\footnote{The catalog associated with this work is available at \citet{Kinson2021_Cat}}. We match every Spitzer~I YSO and YSO candidate from \citet{kinson21} to all objects in our red source catalog using a 1\arcsec{} matching radius. We find that 64 YSOs and YSO candidates from \citet{kinson21} have at least one match in our \emph{JWST} red source catalog. One of these 64 sources corresponds to the background galaxy we show in the rightmost panel of Figure \ref{fig:bkg_gal}. This source was used in the YSO training set in \citet{kinson21}, was identified as a YSO in \citet{jones19}, passed five of our six \emph{JWST} color cuts, and was considered a high-reliability YSO. This illustrates the need for high-resolution imaging to remove background galaxy contaminants. We find that 29 of the 64 sources matched to our red source catalog are classified as YSOs or YSO candidates, 23 are matched to sources that appear to be evolved stars, while the remaining matched sources are unclassifiable due to too few photometric detections or are additional background galaxies. We also find that several of the sources classified as YSOs or YSO candidates in \citet{kinson21} correspond to multiple \emph{JWST} sources within the 1\arcsec{} matching radius. We show an example of this in Figure \ref{fig:mult_match}: The cyan diamond corresponds to the RA and Dec.\ coordinates of the source from the \citet{kinson21} catalog, the teal stars correspond to sources we classify as evolved stars, and the yellow circles are sources that we classify as YSOs or YSO candidates. These five objects from our red source catalog lie within 0.7\arcsec{} of the \citet{kinson21} source and would have all been blended in the previously available UKIRT and \textit{Spitzer} photometry.

\begin{figure*}
    \includegraphics[width=0.32\textwidth]{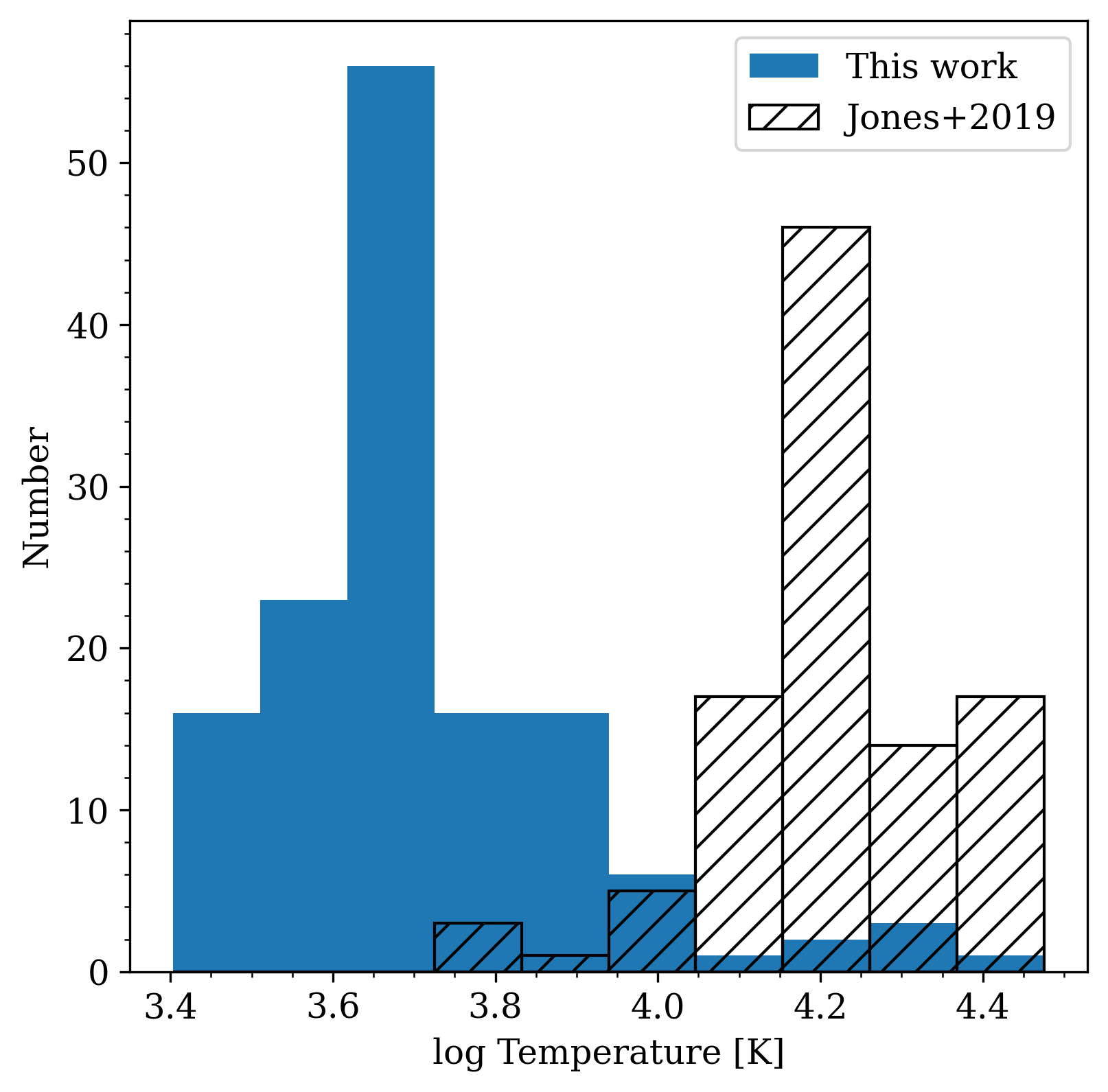}
    \includegraphics[width=0.32\textwidth]{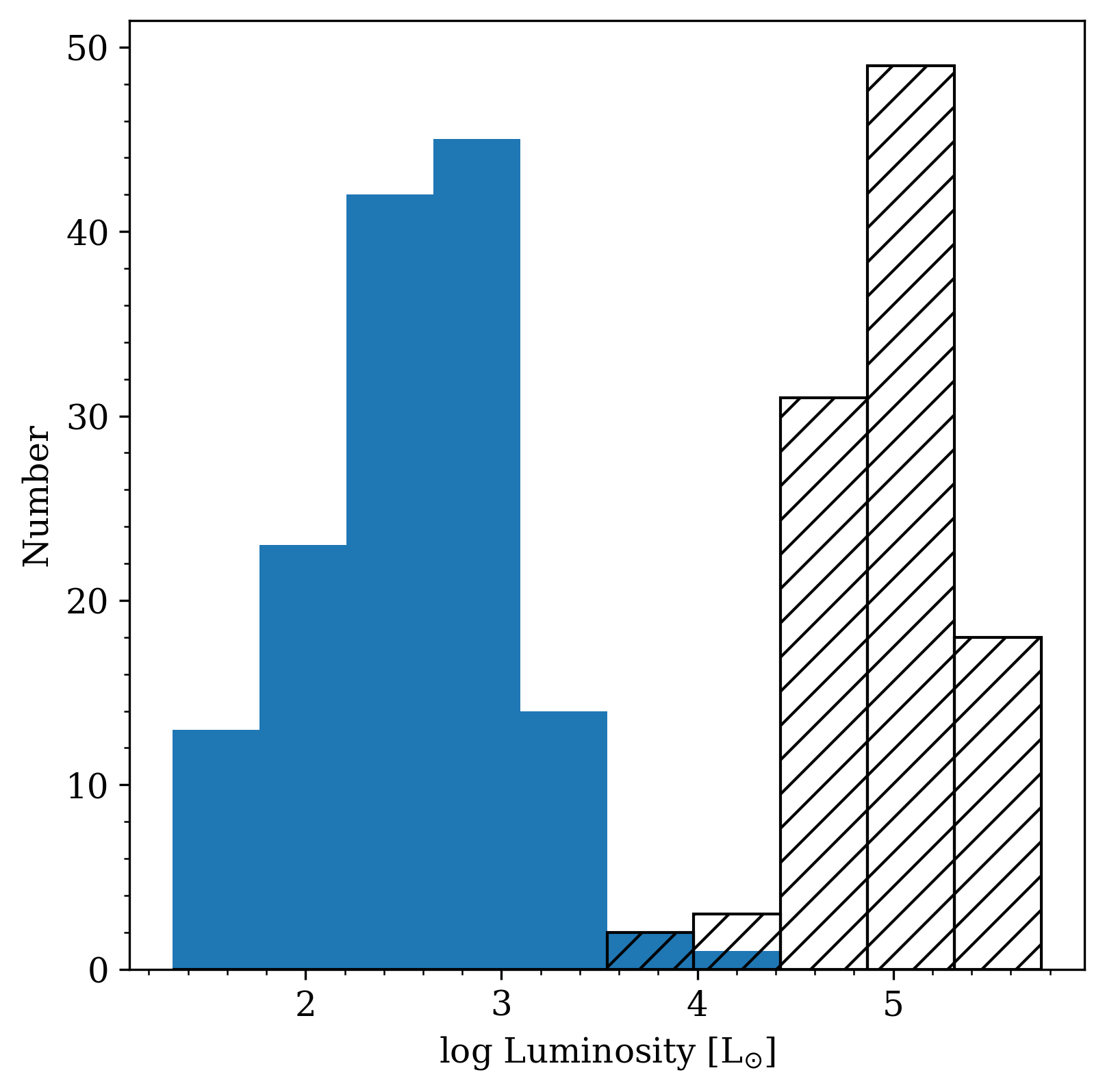}
    \includegraphics[width=0.32\textwidth]{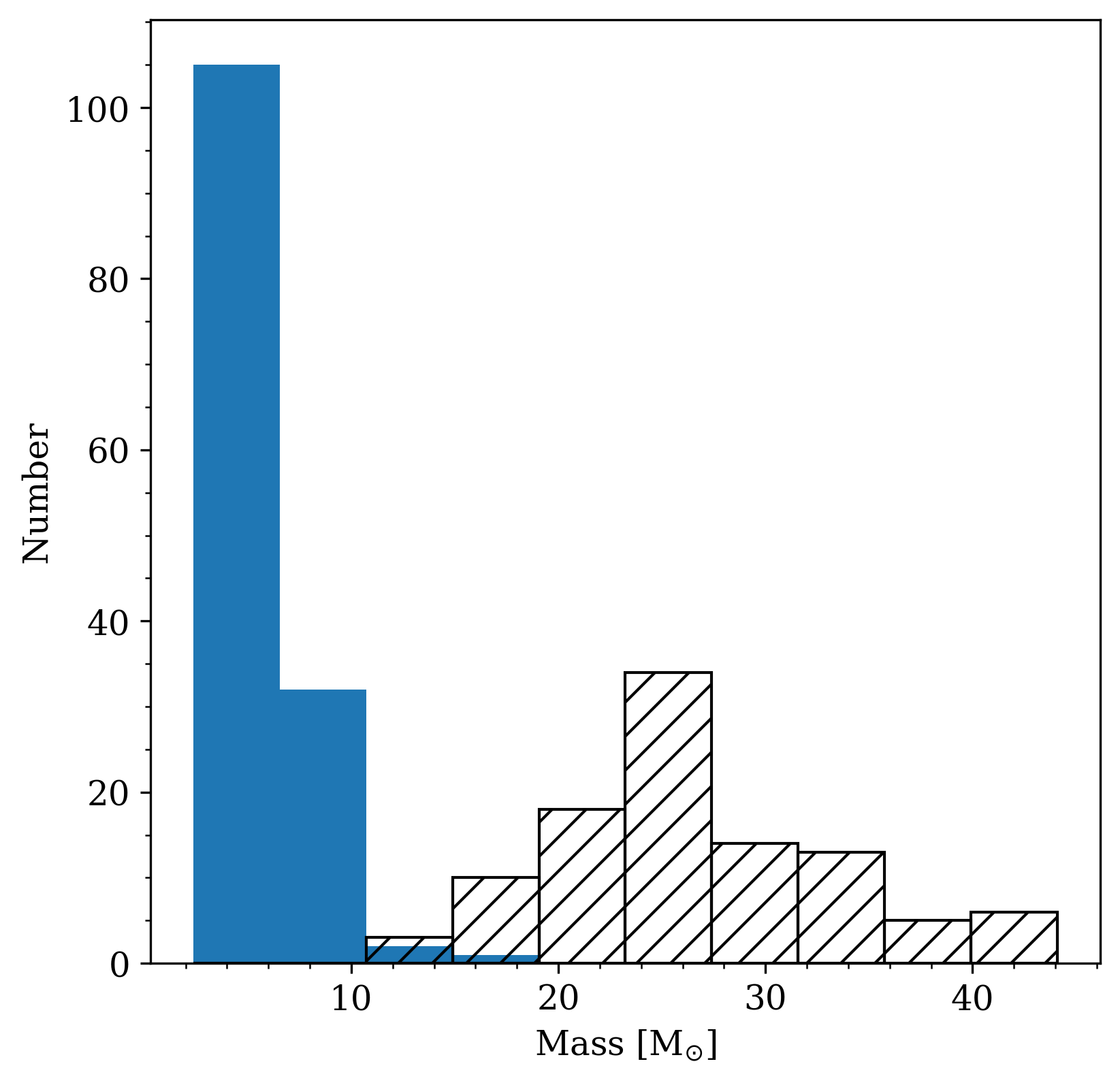}
    \caption{Distribution of temperature (left), luminosity (middle), and mass (right) of the 129 YSOs we identify with \emph{JWST} (blue histograms), compared to the work of \citet{jones19} (hatched histograms). The \emph{JWST}-identified YSOs are overall less massive and less luminous than the \textit{Spitzer}-identified YSOs of \citet{jones19}. This may also reflect limitations due to crowding and blending of sources in the \textit{Spitzer} observations.}
    \label{fig:yso_props}
\end{figure*}

Figure \ref{fig:yso_props} shows how this blending of sources in \textit{Spitzer} observations affects the YSO properties that are subsequently extracted from SED fitting. The blue histograms in Figure \ref{fig:yso_props} show the temperature, luminosity, and mass distribution of YSOs identified in this work, as determined from the best-fit \citet{Robitaille2017} YSO SED models. The black histograms in Figure \ref{fig:yso_props} show the temperature, luminosity, and mass distribution of sources within Spitzer~I from \citet{jones19} determined in the same way. The YSOs in this work have a median temperature of $\mathrm{T} = 4,873${\raisebox{0.5ex}{\tiny$^{+2,443}_{-1,354}$}}~K, median luminosity of $\mathrm{log \: (L/L_{\odot}}) = 2.58${\raisebox{0.5ex}{\tiny$^{+2.84}_{-2.48}$}}, and a median mass of $\mathrm{M} = 5.48${\raisebox{0.5ex}{\tiny$^{+1.86}_{-1.93}$}}~M$_{\odot}$ (where the upper and lower uncertainties correspond to the 84.1 and 15.9 percentiles determined from cumulative distribution functions). The YSO candidates in \citet{jones19} have a median temperature that is $\sim$3.6 times greater ($\mathrm{T} = 17,720${\raisebox{0.5ex}{\tiny$^{+6,480}_{-4,510}$}}~K), median luminosity that is $\sim$200 times greater ($\mathrm{log \: (L/L_{\odot}}) = 4.93${\raisebox{0.5ex}{\tiny$^{+0.39}_{-0.37}$}}), and a median mass that is $\sim$4.7 times greater ($\mathrm{M} = 25.66${\raisebox{0.5ex}{\tiny$^{+7.44}_{-5.49}$}}~M$_{\odot}$) than what is presented in this work. The \textit{Spitzer} data are limited by flux enhancements due to source confusion, because star-forming regions are crowded as is evident in NGC~6822 from the NIRCam three-color image in Figure \ref{fig:color_imgs}. Our results suggest that crowding and source blending are potentially a significant limitation of previous studies.

\begin{figure*}
    \includegraphics[width=\textwidth]{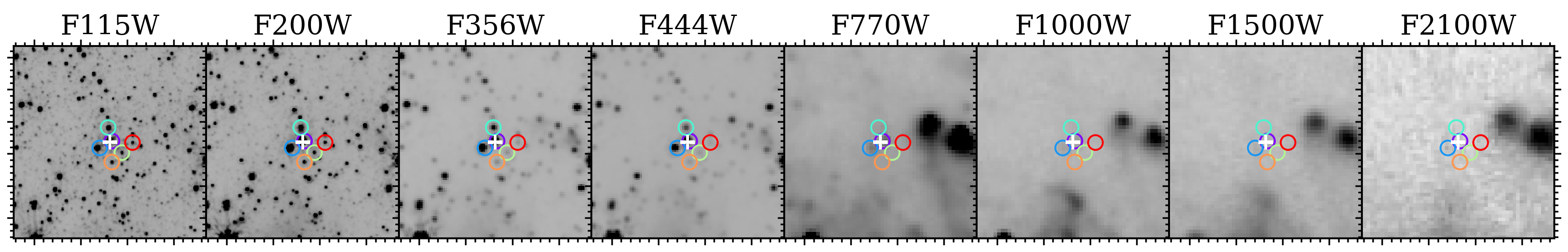}
    \includegraphics[width=\textwidth]{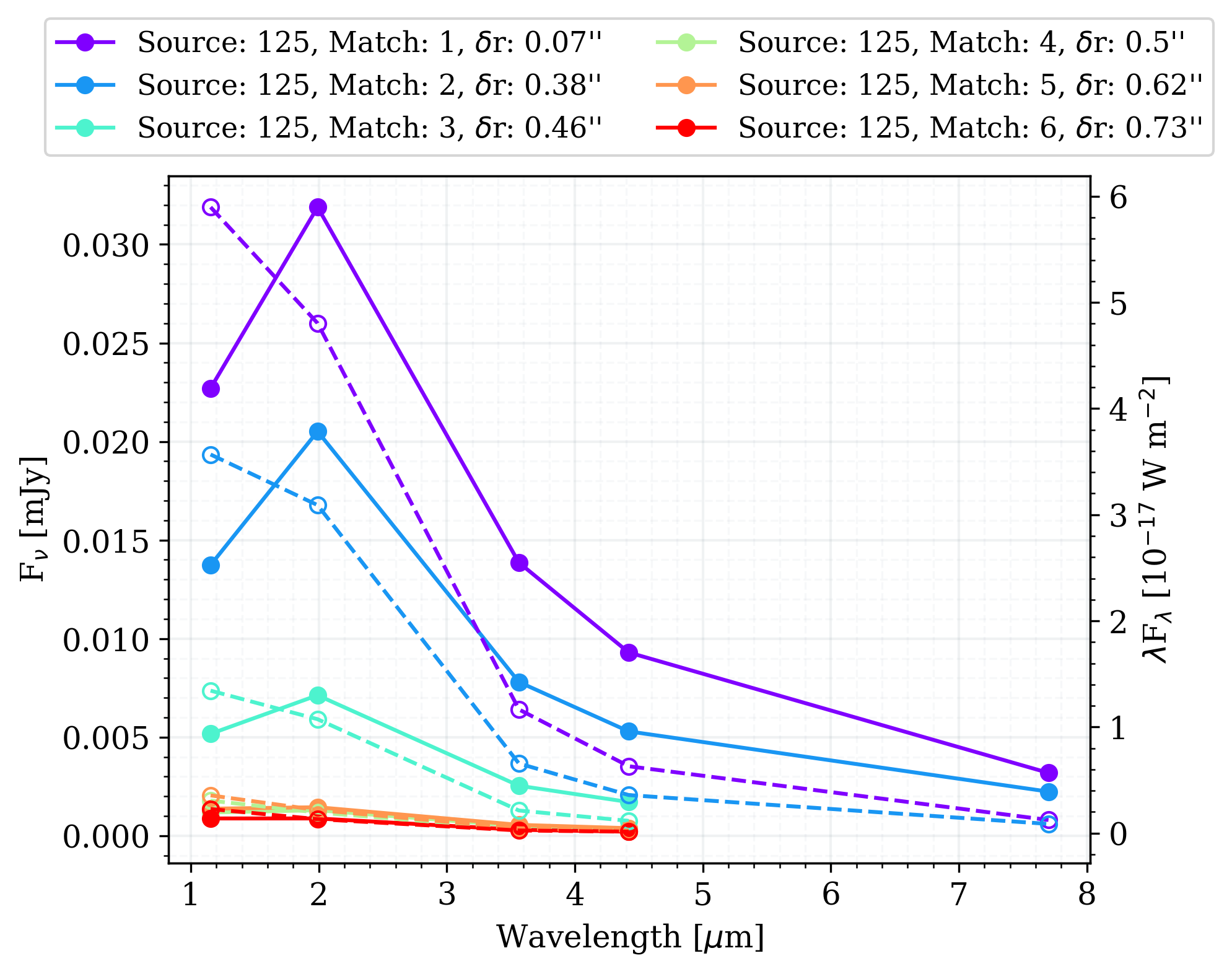}
    \caption{\textit{Top}: NIRCam and MIRI $6\times6$~arcsec$^{2}$ zoom-in images, from shortest to longest wavelength, centered on the coordinates of one YSO from \citet{kinson21} (white cross symbol), that was also identified in \citet{hirschauer20}. The six nearest matches with detections in at least four \emph{JWST} filters, from all sources within Spitzer~I ($\sim$81,000 objects), are indicated with colored circles. \textit{Bottom}: The SEDs of the six nearest matches to the \citet{kinson21} source with flux densities in mJy (left $y-$axis; filled symbols and solid lines) and fluxes in 10$^{-17}$~W\,m$^{-2}$ (right $y-$axis; open symbols and dashed lines). The lines are colored according to distance, in arcseconds, from the \citet{kinson21} source: purple is the closest (0.07\arcsec{}) and red is the farthest (0.73\arcsec{}).}
    \label{fig:k21_check}
\end{figure*}

Next, to investigate why we do not recover a larger fraction of the \citet{kinson21} Spitzer~I YSOs and YSO candidates with our JWST observations, we match the coordinates of each \citet{kinson21} source to all \emph{JWST}-identified sources within Spitzer~I ($\sim$81,000 objects). Subsequently, we inspect the SEDs and images in all eight NIRCam and MIRI bands of the six nearest matches (with detections in at least four \emph{JWST} filters) to each \citet{kinson21} source. We choose to follow up sources with detections in at least four \emph{JWST} filters, because there are often additional close matches that have detections in only one to two filters. We find that sources in the \citet{kinson21} catalog that were identified as YSOs or YSO candidates but have no match to our \emph{JWST} source catalog are bright in NIRCam filters, but are not detected in the mid-IR MIRI filters. We show an example of this in Figure \ref{fig:k21_check}. The bottom panel of this figure shows the SEDs of the six nearest matches to one of the \citet{kinson21} YSOs \citep[source 125 in][]{Kinson2021_Cat} in both flux densities (left $y-$axis; solid lines and filled symbols) and fluxes (right $y-$axis; dashed lines and open symbols). The nearest match (0.07\arcsec{}) is shown in purple and subsequent matches are color-coded according to the legend. The top panel of Figure \ref{fig:k21_check} shows $6\times6$~arcsec$^{2}$ (chosen to match the FWHM of the MIPS 24~\micron{} PSF) zoom-in NIRCam and MIRI images centered on the \citet{kinson21} source, ordered from shortest wavelength (F115W) to longest wavelength (F2100W). The general trend we observe is that there are several sources matched to previously \textit{Spitzer} identified YSOs within 1\arcsec{} that are visible in NIRCam filters, but not in MIRI. In correspondence with this observation, we see that the SEDs of these nearest matches all peak at short wavelengths and diminish at longer wavelengths.

We propose three possibilities for why this may be the case: (1) Sources from \citet{kinson21} that we detect in near-IR wavelengths, but not in the mid-IR, are possibly more evolved YSO candidates. Stage I YSOs are deeply embedded within their parental giant molecular clouds (GMCs) and have a rising spectral index towards the mid- and far-IR \citep{lada87}. In contrast, more evolved Stage II and III YSOs have blown away dust and gas in their vicinity. Therefore, their fluxes are dominated by optical stellar light \citep{lada87, robitaille06} and these YSOs are no longer visible in the mid-IR. Our color cuts may not capture these sources since many of these likely blend in with the general stellar population in the NIRCam CMDs; (2) All sources in common between \citet{jones19} and \citet{kinson21} are best-fit by Stage I YSO models which have rising SEDs toward the mid-IR, and would thus correspond to embedded objects. We see in the top panel of Figure \ref{fig:k21_check} that two bright sources appear to the right from center in all four MIRI filters. It is therefore possible that, due to crowding and resolution limitations, there are mismatched sources across the \emph{JHK}, IRAC, and MIPS photometry that were used to produce the combined catalogs of \citet{jones19}, and that some of these sources are in fact not YSOs. While for the \citet{kinson21} catalog, we see in the NIRCam panels of Figure \ref{fig:k21_check} that there are multiple sources within 0.7\arcsec{} which may blend and confuse color measurements and YSO identification; (3) Finally, the drop in sensitivity at longer wavelengths is a limitation of our \emph{JWST} data that may also contribute to the lack of mid-IR counterparts to the large number of suspected embedded, Stage I YSOs \citep{jones19}. At the most extreme, the faint limit of our NIRCam F115W data is 25.78 magnitudes, while for the MIRI F2100W data it is 14.43 magnitudes \citep[see][]{nally2023}. For our two closest NIRCam and MIRI filters, F444W and F770W, the faint limits are 23.06 and 19.25 magnitudes, respectively. These differences in sensitivity are also reflected in the precipitous drop in source counts toward the mid-IR; within Spitzer~I, the source counts are:\ 56,094 (F115W), 38,599 (F200W), 13,765 (F356W), 11,729 (F444W), 1,145 (F770W), 692 (F1000W), 273 (F1500W), and 72 (F2100W). While we would expect Stage I YSOs to brighten at mid-IR wavelengths, it is possible that they are fainter than what might be expected from MIPS 24~\micron{} if source blending is significant, such as the case we show in the top panel of Figure \ref{fig:k21_check}, where there are two bright sources within the MIPS 24~\micron{} PSF.

\begin{figure}
    \includegraphics[width=\columnwidth]{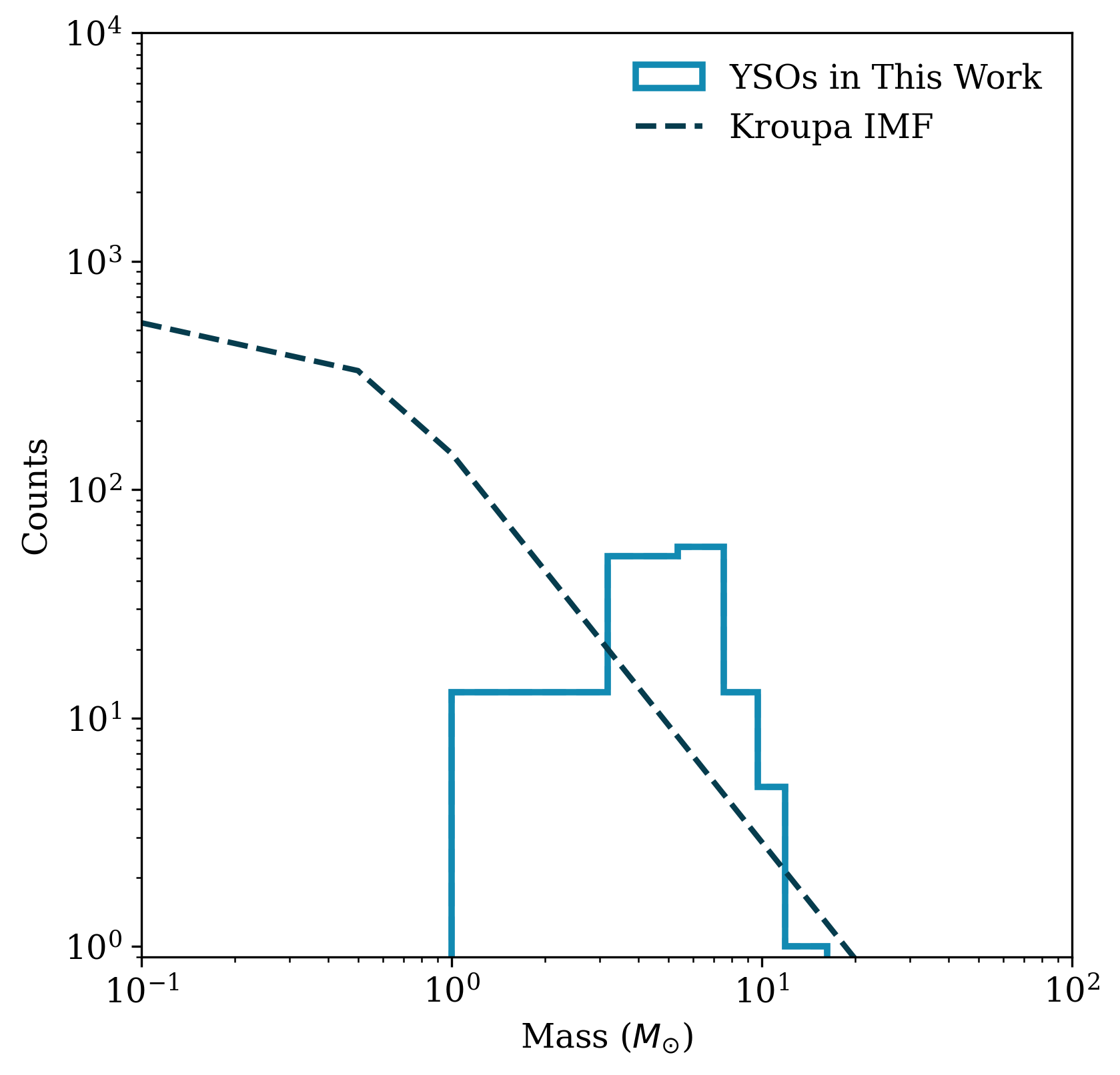}
    \caption{\label{fig:kroupa_imf}Histogram of the 140 YSOs we fit using \citet{Robitaille2017} models (solid line) with the best-fit Kroupa-IMF over-plotted (dashed line). The total stellar mass in Spitzer I derived from best-fit Kroupa IMF is $\mathrm{{\sim}1200\;M_{\odot}}$. Assuming a typical YSO formation timescale of $\mathrm{10^{5}\;yrs}$, the star formation rate of Spitzer I is $\mathrm{0.012\;M_{\odot}\;yr^{-1}}$.}
\end{figure}

\subsection{Comparison to 30 Doradus}
The R136 SSC in 30 Doradus has high stellar densities \citep[$1.5 \times 10^{4}-10^{7}$~M$_{\odot}$\,pc$^{-3}$;][]{selm13} and contains many bright stars with the 30 most massive and luminous having a bolometric luminosity of $\mathrm{4 \times 10^{7}}\;L_{\odot}$ \citep{massey98}. 

The mass range of the 140 \emph{JWST}-identified YSO candidates within Spitzer~I is $\mathrm{2 - 16\;M_{\odot}}$. We fit a three-part Kroupa initial mass function (IMF) to the 129 YSO candidates from this work: $\mathrm{\xi\;\propto\;M^{-0.3}}$ for $\mathrm{M\;\textless\;0.5\;M_{\odot}}$, $\mathrm{\xi\;\propto\;M^{-1.2}}$ for $\mathrm{0.5\;\textless\;M\;\textless\;1.0\;M_{\odot}}$, and $\mathrm{\xi\;\propto\;M^{-1.7}}$ for $\mathrm{M\;\textgreater\;1.0\;M_{\odot}}$ \citep{krou01}. Our inventory of YSO candidates is likely incomplete; estimates of completeness will require further analysis through artificial star testing.
The total stellar mass in Spitzer I derived from best-fit Kroupa IMF is $\mathrm{1200\;M_{\odot}}$ (Figure \ref{fig:kroupa_imf}). Assuming a typical YSO formation timescale of $\mathrm{10^{5}\;yrs}$, the SFR of Sptizer I is $\mathrm{0.012\;M_{\odot}\;yr^{-1}}$.

\citet{nayak23} find the current SFR in 30 Doradus derived from counting the YSO candidates and fitting an IMF to be $\mathrm{0.18\;M_{\odot}\;yr^{-1}}$, roughly an order of magnitude higher than the calculated SFR in Spitzer~I. In contrast, the average SFR over the last 4~Myr in 30 Doradus calculated using H$\mathrm{\alpha}$ is $\mathrm{0.03\;M_{\odot}\;yr^{-1}}$. The SFR in 30 Doradus has increased over time in the last few million years. 
The low SFR of $\mathrm{0.012\;M_{\odot}\;yr^{-1}}$ is consistent with Spitzer~I being a very young and embedded massive cluster, similar to 30 Doradus 4~Myr ago.




\section{Summary and Conclusion} 
\label{sec:summary}
We have presented the first \emph{JWST} NIRCam and MIRI images (GTO PID: 1234; PI: M.\ Meixner) of the central stellar bar of NGC~6822, focused particularly on the Spitzer~I region, a candidate proto-SSC that has been found to host the largest number of YSOs in this galaxy. We aimed to revisit the YSO population of Spitzer~I with this new, high-resolution data. To do so, we have combined NIRCam and MIRI photometry to construct CMDs, and established color selection criteria to identify red sources of interest. We have fit and visually inspected the SEDs of these sources, along with the NIRCam and MIRI images of each, to identify likely YSOs and YSO candidates. We summarize our findings here: \\

\noindent 1. The NIRCam and MIRI data reveal a stark difference in the appearance of Spitzer~I at shorter and longer wavelengths (see Figure \ref{fig:color_imgs}). NIRCam reveals a very densely populated field of stars, with several prominent background galaxies. The F356W observations reveal faint diffuse emission likely associated with the 3.3~\micron{} PAH feature. In contrast, our MIRI observations show a network of dusty filaments traced in 7.7~\micron{} PAH emission. \\

\noindent 2. We establish color selection criteria to pick 1,307 red sources of interest for further follow up and investigation (see Figure \ref{fig:nircam_miri_cmds}). We fit and visually inspect the SEDs of these sources to remove evolved star contaminants. In addition, we visually inspect the images of each source in all eight NIRCam and MIRI filters to identify and remove background galaxy contaminants. This results in a catalog of 140 YSOs and 400 YSO candidates for the Spitzer~I region (see Figure \ref{fig:spatial_dist}). \\

\noindent 3. Matching our catalog to that of \citet{kinson21}, which contains 90 YSOs and 49 YSO candidates within Spitzer~I, we find that 64 of these 139 sources have a match to an object in our catalog. One of these is a prominent background galaxy (see Figure \ref{fig:bkg_gal}), 29 are matched to YSOs or YSO candidates, 23 appear to be evolved stars, and the remaining sources are unclassified or correspond to additional background galaxies. We also find that individual YSOs from the \citet{kinson21} catalog often correspond to multiple \emph{JWST} sources (see Figure \ref{fig:mult_match}). \\

\noindent 4. We also compare the temperatures, luminosities, and masses of the \emph{JWST}-identified YSOs to the \textit{Spitzer}-identified YSOs of \citet{jones19}, extracted from SED fitting (see Figure \ref{fig:yso_props}). We find that these earlier studies over-estimate these properties, which is likely due to the limitations of the \textit{Spitzer Space Telescope} resolution and the blending of multiple sources in a crowded field. \\

\noindent 5. To investigate we do not recover a larger fraction of YSOs from \citet{kinson21} with our \emph{JWST} data, we match the \citet{kinson21} objects to all sources we extract within the Spitzer~I region. We visually inspect the SEDs and images across all filters for the nearest six matches that have detections in at least four \emph{JWST} filters (see Figure \ref{fig:k21_check}). We find that in nearly all cases where there are no \emph{JWST} matches, all nearest matched stars have SEDs that peak at short wavelengths in NIRCam and fall off at longer wavelengths in MIRI. We speculate that the reasons for this might be that: (1) These sources might actually be Stage II or III YSOs which may be faint in MIRI and are not selected by our color cuts; (2) In many cases, we see that within 6\arcsec{} (the FWHM of the \textit{Spitzer} MIPS 24~\micron{} filter), there are nearby sources that appear bright at MIRI and could feasibly have been MIPS sources that were mismatched to IRAC sources in earlier works, and are thus not YSOs in the \citet{jones19} catalog, while for the \citet{kinson21} catalog, there are multiple sources even within 0.7\arcsec{} that may blend and confuse color measurements; and (3) The sensitivity of MIRI drops dramatically at the longest MIRI wavelengths compared to the shortest NIRCam wavelengths, and we may thus be missing sources. 

This work relies on PSF photometry for NIRCam and aperture photometry for MIRI. As work continues to improve the \emph{JWST} calibrations and simulated PSFs, future work will include conducting PSF photometry for our MIRI observations and perhaps extracting fainter sources and increasing our source counts in the mid-IR. As we refine our photometry and source catalogs, we will revisit the YSO population of Spitzer~I and reassess these results and how our \emph{JWST} findings compare to previous studies, and investigate additional statistical tools for identifying and separating various stellar populations. In conclusion, \emph{JWST} has revealed a striking new view of the central stellar bar of NGC~6822 and of the Spitzer~I star-forming region, and it shows that its high resolution is necessary to properly identify individual YSOs, characterize their properties, and remove background contaminants. 

\section*{Acknowledgements}
This work is based on observations made with the NASA/ESA/CSA James Webb Space Telescope. The data were obtained from the Mikulski Archive for Space Telescopes at the Space Telescope Science Institute, which is operated by the Association of Universities for Research in Astronomy, Inc., under NASA contract NAS 5-03127 for \emph{JWST}. These observations are associated with program \#1234. All of the data presented in this paper were obtained from the Mikulski Archive for Space Telescopes (MAST) at the Space Telescope Science Institute. The specific observations analyzed can be accessed via \dataset[doi:10.17909/3s1j-gp52]{https://doi.org/10.17909/3s1j-gp52}.

LL acknowledges support from the NSF through grant 2054178.
OCJ acknowledges support from an STFC Webb fellowship. 
CN acknowledges the support of an STFC studentship.
MM and NH acknowledge that a portion of their research was carried out at the Jet Propulsion Laboratory, California Institute of Technology, under a contract with the National Aeronautics and Space Administration (80NM0018D0004).
PJK acknowledges support from the Science Foundation Ireland/Irish Research Council Pathway programme under Grant Number 21/PATH-S/9360.
ON acknowledges the NASA Postdoctoral Program at NASA Goddard Space Flight Center, administered by Oak Ridge Associated Universities under contract with NASA.
ASH is supported in part by an STScI Postdoctoral Fellowship.

\vspace{5mm}
\facilities{\emph{JWST}(MIRI, NIRCam)}

\software{aplpy \citep{aplpy2019},
          astropy \citep{astropy13,astropy2018,astropy2022},
          image1overf.py \citep{1fcor},
          JHAT \citep{jhat},
          SED Fitter \citep{Robitaille2007},
          \textsc{starbugii} \citep{nally22}.
          \textsc{topcat} \citep{Taylor2005}}



\bibliography{references}{}
\bibliographystyle{aasjournal}

\end{document}